\documentclass[aps,pre,onecolumn]{revtex4}
\usepackage{graphicx, bm, bbm,amsmath, amssymb, epsfig}

\newcommand{\nn}{\noindent}

\newcommand{\bq}{\begin{align}}
\newcommand{\eq}{\end{align}}

\begin{document}
\title{Intermittent sliding locomotion of a two-link body}
\author{Silas Alben$^*$ and Connor Puritz}
\affiliation{Department of Mathematics, University of Michigan,
Ann Arbor, MI 48109, USA}
\email{alben@umich.edu}

\date{\today}

\begin{abstract}
We study the possibility of efficient intermittent locomotion for two-link bodies that slide by changing their interlink angle periodically in time. We find that the anisotropy ratio of the sliding friction coefficients is a key parameter, while solutions have a simple scaling dependence on the friction coefficients' magnitudes.  With very anisotropic friction, efficient motions
involve coasting in low-drag states, with rapid and asymmetric power and recovery strokes. As the anisotropy decreases, burst-and-coast motions change to motions with long power strokes and short recovery strokes, and roughly constant interlink angle velocity on each. These motions are seen in the spaces of
sinusoidal and power-law motions described by two and five parameters, respectively. Allowing the duty cycle to vary greatly increases the motions' efficiency compared to the case of symmetric power and recovery strokes. Allowing
further variations in the concavity
of the power and recovery strokes only improves the efficiency further when friction is very anisotropic. 
Near isotropic friction, a variety of optimally efficient motions are found with more complex waveforms. Many of the optimal sinusoidal and power-law motions are similar to those that we find with an optimization search in the space of more general periodic functions (truncated Fourier series). When we increase the resistive force's power law dependence on velocity, the optimal motions become smoother, slower, and less efficient, particularly near isotropic friction.  
\end{abstract}

\pacs{}

\maketitle

\section{Introduction \label{sec:Introduction}}

In this paper we study the optimal kinematics for intermittent locomotion by simple (two-link) bodies sliding
on rough surfaces. Intermittent locomotion occurs when propulsive forces are 
applied (very) nonuniformly in time, perhaps for only a brief interval \cite{kramer2001behavioral}. 
Such locomotion is often divided into
two phases, termed burst and coast, power and recovery, or thrust and drag, for example. 
Much previous work has studied the optimal kinematics of bodies swimming
in fluids, at low, intermediate ($O(1)$), and high Reynolds numbers (i.e.~inverse viscosities).
Bodies in fluids experience velocity-dependent drag forces that can penalize more rapid and
intermittent motions to some degree. At low Reynolds number, for example, it can be shown that 
the most efficient swimming motions exert constant mechanical power \cite{BeKoSt2003a}, or have a constant stroke speed 
\cite{AvGaKe2004a}. At higher Reynolds numbers, fluid drag typically scales as velocity squared, and
steadier swimming speeds and propulsive forces may be more efficient for drag-based locomotion \cite{alben2010coordination,de2013don,larson2014effect,ford2019hydrodynamics}. Interestingly, for undulatory high-Reynolds-number swimmers, intermittent (e.g.~``burst-and-coast") swimming can be more efficient than steady swimming \cite{mchenry2005mechanical}, in part because of
boundary layer thinning during steady swimming \cite{lighthill1971large,weihs1974energetic} and differences in 
vortex shedding dynamics \cite{akoz2018unsteady,akoz2019large}. Jellyfish, octopus, and scallops use
intermittent, jet-propelled swimming, and there is controversy about its efficiency relative to steady undulatory swimming \cite{daniel1985cost,o1991invertebrate,anderson2000mechanics,
alben2013efficient,gemmell2014exploring,park2015dynamics,hoover2017quantifying,gemmell2018widespread,hoover2019pump}.

For bodies sliding on a dry surface, the simplest force model---the Coulomb frictional force---is independent of velocity magnitude. Therefore, there is the possibility of a smaller
energy penalty for large, fluctuating velocities in this case.  In the natural world, snake locomotion
presents many examples of sliding locomotion \cite{gray1946mechanism,jayne1988mechanical,hirosebiologically,MaHu2012a,lillywhite2014snakes,marvi2014sidewinding}. Many familiar gaits for steady snake locomotion 
(e.g.~lateral undulation, sidewinding, rectilinear motion) 
involve waves of body deformation that can propel the snake at nearly constant speed \cite{lillywhite2014snakes}. However,
snake locomotion can be more intermittent and involve large accelerations, particularly when the substrate has
low friction or the snake is trying to escape a predator.  For example, in ``slide pushing," the snake moves with vigorous and irregular undulations that propagate backward along the body, propelling the body forward by sliding friction \cite{gans1984slide,lillywhite2014snakes}. Thus far, most theoretical studies of sliding locomotion have focused on cases where body acceleration is negligible \cite{GuMa2008a,HuNiScSh2009a,HaCh2011a,HuSh2012a,AlbenSnake2013,wang2014optimizing,bittner2018geometrically}, with \cite{wang2018dynamics} as an exception. The goal of the present work is to determine some of the basic physical mechanisms that result in intermittent sliding locomotion, and to what extent it can be efficient. We study perhaps the simplest case---a two-link body. The advantage of this
system is that the body shape has a single degree of freedom, as does the translational motion, due to the (assumed) bilateral symmetry. This simplification allows us to describe more completely how dynamics depend on physical effects. We focus on three: 
the role of body inertia, frictional anisotropy, and the resistive force law. 
Even for this simple body, the motion has a complex, nonlinear dependence on the body shape kinematics. For a two-link body with bilateral
symmetry, intermittent locomotion is the only possibility, which removes any question about the type of locomotion being studied.





\section{Model \label{sec:Model}}

\begin{figure} [h]
           \begin{center}
           \begin{tabular}{c}
               \includegraphics[width=4in]{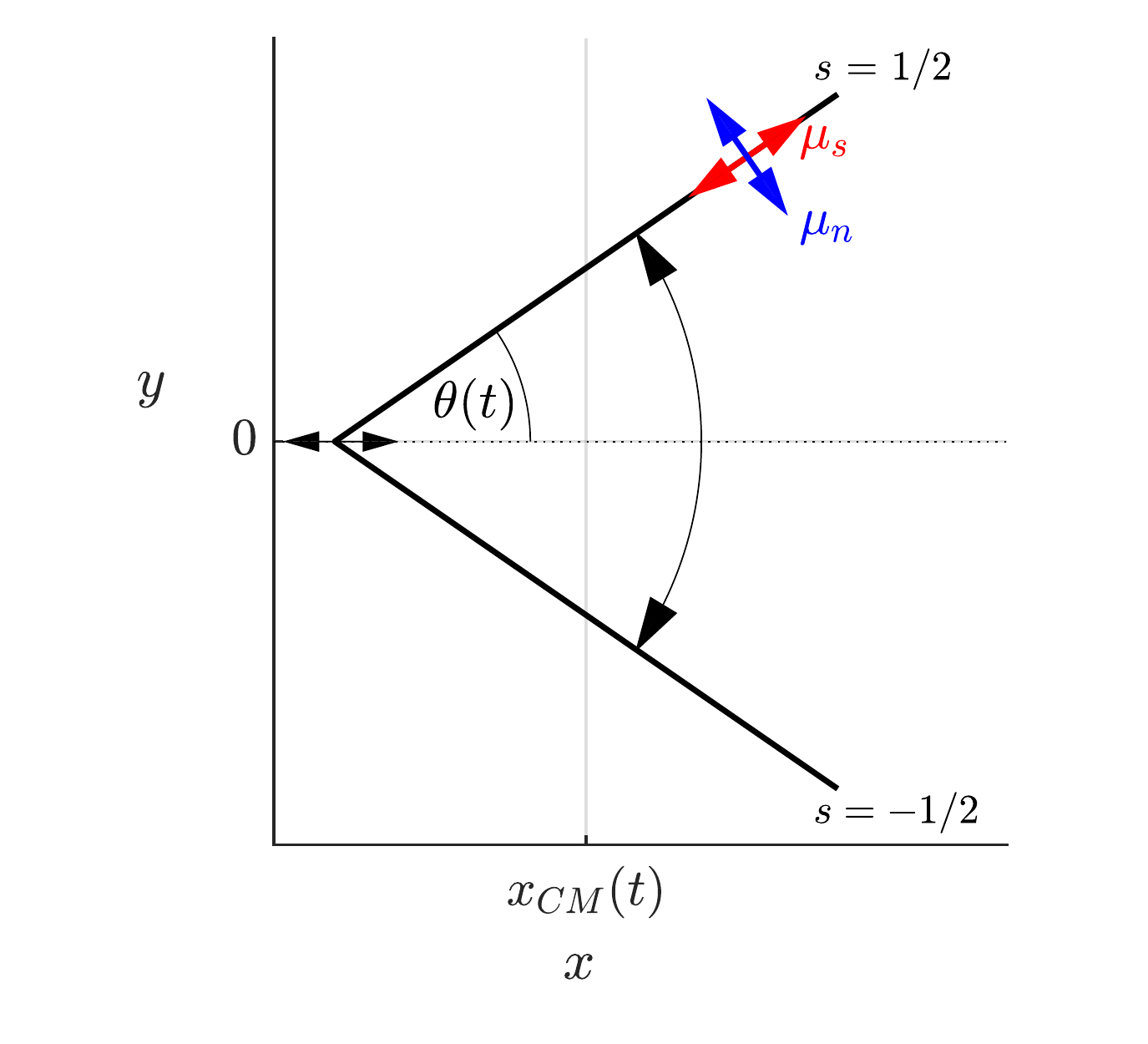}\\
           \vspace{-.25in}
           \end{tabular}
          \caption{\footnotesize Schematic diagram of a two-link
body parametrized by arc length $s$ (nondimensionalized by body length),
 at an instant in time.  The tangent angle on the upper half 
(equal to one-half the opening angle) is prescribed as
$\theta(t)$.
 Vectors representing
tangential and normal velocities are shown with
the corresponding friction coefficients $\mu_s$ and $\mu_n$.
 \label{fig:ScallopSchematic}}
           \end{center}
         \vspace{-.10in}
        \end{figure}

A schematic
diagram of the body is shown in figure \ref{fig:ScallopSchematic}.
We denote half the opening angle
by $\theta(t)$, and it is prescribed as a control variable, periodic in time. 
The position of the body is 
$\mathbf{X}(s,t) = (x(s,t), y(s,t)), -L/2 \leq s \leq L/2$, and 
as $\theta(t)$
varies, the body slides in the plane with three degrees of freedom, two translational and one rotational. 
These are determined by three equations, Newton's laws for the rates of change of the body's total linear
and angular momenta:
\begin{align}
\int_{-L/2}^{L/2} \rho \partial_{tt} x \,ds &=\int_{-L/2}^{L/2} f_{x} \,ds, \label{fx0} \\
\int_{-L/2}^{L/2} \rho \partial_{tt} y \,ds &=\int_{-L/2}^{L/2} f_{y} \,ds, \label{fy0} \\
\int_{-L/2}^{L/2}  \rho \mathbf{X}^\perp \cdot \partial_{tt} \mathbf{X} \,ds
&=\int_{-L/2}^{L/2} \mathbf{X}^\perp \cdot \mathbf{f}\,ds, \label{torque0}
\end{align}
\nn where $\rho$ is the body mass per unit length (assumed uniform) and
 $\mathbf{f}$ is the frictional force acting on the body,
\begin{align}
\mathbf{f}(s,t) &\equiv -\rho g \mu_n
\left( \widehat{\partial_t{\mathbf{X}}}_\delta\cdot \hat{\mathbf{n}} \right)\hat{\mathbf{n}}
-\rho g \mu_s \left( \widehat{\partial_t{\mathbf{X}}}_\delta\cdot \hat{\mathbf{s}} \right)\hat{\mathbf{s}}, \label{frictiondelta} \\
\widehat{\partial_t{\mathbf{X}}}_\delta &\equiv \frac{\left(\partial_t x, \partial_t y\right)}{\sqrt{\partial_t x^2 +\partial_t y^2 + \delta^2}}. \label{delta}
\end{align}
\noindent Here $g$ is gravitational acceleration, and $\hat{\mathbf{s}} = (\mbox{sgn}(s)\cos{\theta(t)}, \sin{\theta(t)})$ 
and $\hat{\mathbf{n}} = (-\sin{\theta(t)}, \mbox{sgn}(s)\cos{\theta(t)})$ are the unit tangent and normal vectors
along the body, respectively. $\mathbf{f}$ is a spatially-distributed Coulomb kinetic frictional
force, used by \cite{GuMa2008a,HuNiScSh2009a,HuSh2012a,JiAl2013} and other
recent works to model the force of the substrate on the sliding body. Different friction coefficients $\mu_n$ and $\mu_s$ apply for motions
in the normal and tangential directions, respectively. Their difference is due to anisotropy of the body surface, e.g.~due to scales (for snakes)
or wheels (for snake robots \cite{hirosebiologically,HaCh2010b,astley2015modulation,aguilar2016review}). For common body and substrate materials, $\mu_n$ and $\mu_s$ are usually less than unity \cite{persson2000sliding}. A drag law somewhat similar to (\ref{frictiondelta}) has also been used to describe forces in dry granular media \cite{ding2012mechanics,aguilar2016review}.
In (\ref{frictiondelta}), $\widehat{\partial_t{\mathbf{X}}}_\delta$ is the normalized
velocity with a small regularization parameter $\delta$ (typically $10^{-5}$ here), that smoothes the discontinuity
at zero velocity. In \cite{alben2019efficient} we found that nonzero $\delta$ allows (\ref{fx0})-(\ref{torque0}) to be solved for certain types of body motions for which no solution
exists with $\delta = 0$. Similar
types of Coulomb friction regularization (sometimes involving the arctangent function) have also been used in 
dynamical simulations involving friction \cite{popov2010contact,pennestri2016review,yona2019wheeled}. Many
regularizations (including ours) involve a frictional force that rises monotonically from zero at zero velocity to the kinetic friction
force \cite{quinn2004new,borello2010comparison,ligier2015few,gholami2016linear}, while others (e.g.~\cite{vigue2017regularized}) allow for a 
nonmonotonic behavior near zero velocity, to simulate the effect of a static
friction coefficient that is greater than the kinetic friction coefficient, and allow for stick-slip transitions. In many works (including the present one), variations in
regularization cause only small differences in the body's trajectory and the work it does on the substrate through friction \cite{popov2010contact,pennestri2016review}, which are the focus of this work. In particular, we find very little
change in the body's motion and rate of work done on the ground through friction when $\delta \lesssim 10^{-2}$. 

If the body has zero net angular momentum, the frictional force distribution (\ref{frictiondelta})
is symmetric about the body's line of symmetry ($y = 0$ in figure \ref{fig:ScallopSchematic}), so
the net frictional force lies along this line and there is zero net torque from friction. Thus, if there is no angular momentum
initially, there is no angular momentum going forward, and equations (\ref{fx0})--(\ref{torque0})
reduce to a single equation for the translational motion along the line of symmetry, given by (\ref{fx0}) for the
orientation chosen here (i.e.~in figure \ref{fig:ScallopSchematic}). Thus the motion of the body is described by a scalar
function of time, the $x$-center-of-mass $x_{CM}(t)$ shown in figure \ref{fig:ScallopSchematic}. We mostly focus on its time-derivative, the spatial average of the body's speed.

Our goal is to determine body kinematics
(i.e.~$\theta(t)$) that result in efficient locomotion, in terms of relevant dimensionless parameters.
To do so,
we nondimensionalize all variables in the dynamical equation (\ref{fx0}) by scaling length by
the body length $L$, and time by an intrinsic time scale $\sqrt{L/g}$. 
We use $\rho g L^2$ as the intrinsic scale for energy, i.e.~to nondimensionalize the work done against friction. 
Next, we write the position of the body in terms of $\theta(t)$ and $x_{CM}(t)$. 
On $0 \leq s \leq 1/2$, (the upper half in figure \ref{fig:ScallopSchematic}), 
\begin{align}
x(s,t) &= x_{CM}(t) + (s - 1/4) \cos{\theta(t)} \;, \; y(s,t) = s  \sin{\theta(t)}   \;, \; 0 \leq s \leq 1/2,
\end{align}
\noindent and the lower half, on $-1/2 \leq s \leq 0$, is given by symmetry:
\begin{align}
x(s,t) &= x(-s, t)  \;, \; y(s,t) = -y(-s, t)\;, \; -1/2 \leq s \leq 1/2. \label{symm}
\end{align}


The dimensionless equation (\ref{fx0}) is 
\begin{align}
\ddot{x}_{CM}(t) =\int_{-1/2}^{1/2} f_{x}(s,t) \,ds = 2 \int_0^{1/2} f_{x}(s,t) \,ds \quad ; \quad
f_{x}(s,t) = -\mu_n
\left( \widehat{\partial_t{\mathbf{X}}}_\delta\cdot \hat{\mathbf{n}} \right)n_x
-\mu_s \left( \widehat{\partial_t{\mathbf{X}}}_\delta\cdot \hat{\mathbf{s}} \right)s_x. \label{Newton}
\end{align}
\noindent Given $\theta(t)$, the problem is reduced to solving (\ref{Newton}),
a nonlinear ODE for $\dot{x}_{CM}(t)$ (which appears nonlinearly in $f_{x}$). We consider $\theta(t)$ that are 
periodic with dimensionless period $T$. 
For such $\theta(t)$,  $\dot{x}_{CM}(t)$ also becomes periodic with period $T$ after an initial transient 
motion; our focus is the eventual periodic steady state. It is useful to solve (\ref{Newton}) on a time domain
that is fixed for all choices of the period $T$. Thus 
we define $\tau$ to be a new time variable scaled by $T$, and write $\theta(t)$ and $\dot{x}_{CM}(t)$ as
functions of $\tau$:
\begin{align}
\theta(t) = \phi(\tau(t)) \quad ; \quad \dot{x}_{CM}(t) = \frac{1}{T}u(\tau(t)) \quad ; \quad \tau(t) \equiv t/T \quad ; \quad \delta_\tau \equiv T\delta \label{tau}
\end{align}
\nn Rewriting equation (\ref{Newton}) in terms of $u$ and $\tau$ (with $\phi$ and $\delta_\tau$ in place of $\theta$ and $\delta$, respectively, in $f_{x}$), we obtain
\begin{align}
\frac{1}{T^2}\dot{u}(\tau) = 2 \int_0^{1/2} f_{x}(s,\tau) \,ds. \label{Newtonu}
\end{align}
\nn We compute solutions $u(\tau)$ that are periodic on a $\tau$-interval of length unity. The period $T$ appears only in the
factor on the left side of (\ref{Newtonu}), where it sets the magnitude of the inertia term. Considering the simple dependence of $f_{x}$ on $\mu_n$ and $\mu_s$ we see that solutions $u(\tau)$ depend on $T$, $\mu_n$, and $\mu_s$ only through the two combinations $T^2\mu_s$ and
$T^2\mu_n$. After solving for $u(\tau)$, the true body speed $\dot{x}_{CM}(t)$ is obtained via (\ref{tau}).

Given the body kinematics ($T$ and $\phi(\tau)$), we solve for the motion of the body $u(\tau)$ with two different 
methods, either of which may be faster depending on the values of $T^2\mu_s$ and
$T^2\mu_n$ and the number of functions $\phi(\tau)$ being considered. The first method 
integrates (\ref{Newtonu}) forward in time using a 2nd-order Runge Kutta method. The body starts from rest ($u(0) = 0$), and the computation continues until $u(\tau)$ converges to a periodic solution. The second method discretizes (\ref{Newtonu})
as a boundary value problem with periodic boundary conditions, yielding a coupled system of equations for $u$ on a uniform grid of values of $\tau \in [0, 1)$. Instead of a nonlinear Newton-type iteration, a semi-implicit linearized iteration
based on that in \cite{alben2019efficient} is used, with a geometric rate of convergence in most cases.
Both methods use a trapezoidal rule for the
integral in (\ref{Newtonu}). 

Instead of using $T^2\mu_s$ and $T^2\mu_n$ as the control parameters, we will explain that it is preferable to use their ratio, $\mu_n/\mu_s$, and their product raised to the 1/4 power, $(\mu_s\mu_n)^{1/4}T$. The ratio measures the frictional anisotropy, and we will show that it is the key parameter governing the behavior of the solutions. When $\mu_s$ and $\mu_n$ are fixed, varying $(\mu_s\mu_n)^{1/4}T$ is equivalent to varying the period of the motion. Both
$\mu_n/\mu_s$ and $(\mu_s\mu_n)^{1/4}T$ can have a strong effect on the efficiency of the body motions.


The first main output quantity of interest is the time-averaged speed,
\begin{align}
U \equiv \lim_{t \to \infty} \frac{1}{T}\int_t^{t+T} \dot{x}_{CM}(t') \,dt' =  \frac{1}{T} \lim_{\tau \to \infty} \int_\tau^{\tau+1} u(\tau') \,d\tau'. \label{U}
\end{align}
\nn Although $u$ depends on $T$, $\mu_n$, and $\mu_s$ only through two combinations, e.g.~$\mu_n/\mu_s$ and $T(\mu_s\mu_n)^{1/4}$, $U$ depends on the three parameters
$T$, $\mu_n$, and $\mu_s$ independently because of the $1/T$ prefactor of the rightmost limit in (\ref{U}). However, $(\mu_s\mu_n)^{-1/4} U$ depends only on $\mu_n/\mu_s$ and $T(\mu_s\mu_n)^{1/4}$:
\begin{align}
(\mu_s\mu_n)^{-1/4} U =  \frac{1}{(\mu_s\mu_n)^{1/4}T} \lim_{\tau \to \infty} \int_\tau^{\tau+1} u(\tau') \,d\tau'. \label{Uscale}
\end{align}
\nn We use this expression as follows. For a given $\mu_n/\mu_s$ value, one can find the maximum of $(\mu_s\mu_n)^{-1/4} U$ with respect to $T(\mu_s\mu_n)^{1/4}$ and the other kinematic parameters defining $\phi(\tau)$. 
Hence the speed has only a simple scaling dependence on the product $\mu_n\mu_s$. In particular, it only affects the optimal kinematics for speed by rescaling the period of the optimal motion. 

The second main quantity of interest is the time-averaged power (rate of work) done by the body against friction,
\begin{align}
P \equiv \lim_{t \to \infty}\frac{1}{T}\int_t^{t+T} \int_{-1/2}^{1/2} -\mathbf{f}(s,t') \cdot \partial_t \mathbf{X}(s,t') \,ds \,dt'= \lim_{t \to \infty} \frac{1}{T}\int_t^{t+T} 2 \int_{0}^{1/2} \frac{\mu_n
\left(\partial_t\mathbf{X}\cdot \hat{\mathbf{n}} \right)^2
+ \mu_s \left( \partial_t\mathbf{X}\cdot \hat{\mathbf{s}} \right)^2}{\sqrt{\partial_t x^2 +\partial_t y^2 + \delta^2}} \,ds \,dt'. \label{power}
\end{align}
\nn When all the terms in (\ref{power}) are written in terms of $u(\tau), \phi(\tau)$, and $\delta_\tau$,
we find that $P$ is $1/T^3$ times an integral involving these terms and the parameters $\mu_n/\mu_s$ and $T(\mu_s\mu_n)^{1/4}$. Therefore $(\mu_s\mu_n)^{-3/4} P$ is $((\mu_s\mu_n)^{1/4}T)^{-3}$ times the same integral, so 
$(\mu_s\mu_n)^{-3/4} P$, like $(\mu_s\mu_n)^{-1/4} U$, depends only on $\mu_n/\mu_s$, $T(\mu_s\mu_n)^{1/4}$, and the other kinematic parameters defining $\phi(\tau)$.

An efficiency can be defined as
\begin{align}
\lambda \equiv U/P \label{lambda}
\end{align}
\nn as in \cite{HuNiScSh2009a,HuSh2012a,AlbenSnake2013,JiAl2013}. These works considered the limit
$T \to \infty$, in which case $P$ and $U$ are proportional for a given periodic motion, independent of
the parametrization of time. For finite $T$,  $P$ and $U$ are only approximately 
proportional as $T$ changes. One could determine the motions that maximize $U$ separately for each fixed value of $P$, but for simplicity, we continue to use $\lambda$ as a measure of efficiency, and maximize its magnitude over motions with various $P$ considered together. 

Let $\mu_{min} \equiv$ min($\mu_n,\mu_s$). In the limit $\delta \to 0$,
\begin{align}
P  \geq   \lim_{t \to \infty} \frac{1}{T}\int_t^{t+T} \int_{-1/2}^{1/2} \mu_{min} \| \partial_t \mathbf{X}(s,t') \| \,ds \,dt' \geq \mu_{min}   \lim_{t \to \infty} \left\Vert \frac{1}{T}\int_t^{t+T}   \int_{-1/2}^{1/2}  \partial_t \mathbf{X}(s,t')  \,ds \,dt'  \right\Vert
= \mu_{min}  U ,
\end{align}
so $\lambda$ has an upper bound $\lambda_{ub} \equiv 1/\mu_{min}$ in the limit $\delta \to 0$. This is the efficiency for a unidirectional motion, in the direction of minimum friction. We will examine the scaled efficiency 
\begin{align}
\sqrt{\mu_s\mu_n}\,\lambda = \frac{(\mu_s\mu_n)^{-1/4}U}{(\mu_s\mu_n)^{-3/4}P}
\end{align}
\nn which, following the previous discussion, depends only on $\mu_n/\mu_s$, $(\mu_s\mu_n)^{1/4}T$, and the other kinematic parameters defining $\phi(\tau)$, as does the relative efficiency
\begin{align}
\frac{\lambda}{\lambda_{ub}} = \frac{\sqrt{\mu_s\mu_n}\,\lambda}{\sqrt{\mu_s\mu_n}/\mu_{min}} = \frac{\sqrt{\mu_s\mu_n}\,\lambda}{\mbox{max}(\sqrt{\mu_n/\mu_s}, \sqrt{\mu_s/\mu_n})}. \label{relefficiency}
\end{align}

\begin{figure} [h]
           \begin{center}
           \begin{tabular}{c}
               \includegraphics[width=6.5in]{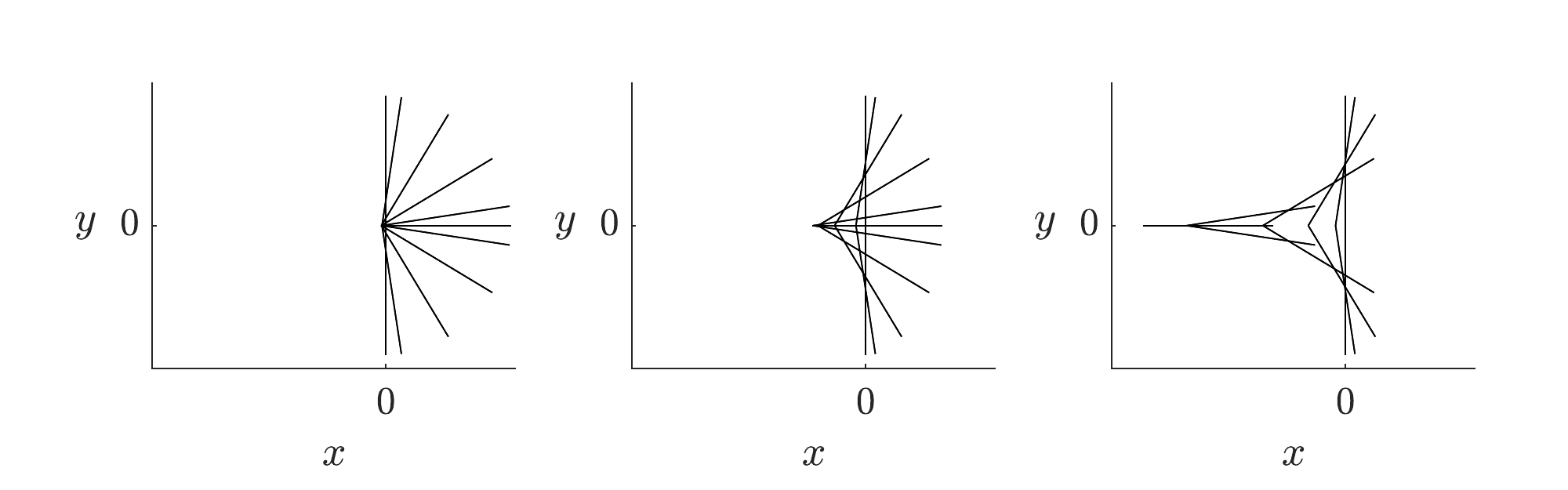}\\
           \vspace{-.25in}
           \end{tabular}
          \caption{\footnotesize  Oscillatory motion of
a body in the limit of very slow movements (periodic with period $T \to \infty$) with friction
coefficient ratio $\mu_n/\mu_s$ equal to $2^{-7}$ (left), $2^0$ (center), and $2^7$ (right).
 \label{fig:TauInfFigure}}
           \end{center}
         \vspace{-.10in}
        \end{figure}

For any kinematic profile $\phi(\tau)$, we can show that $U \to 0$ but $P$ generally remains nonzero in the limit $T \to \infty$, 
so two-link bodies do not locomote in this limit. Here the problem depends only on $\mu_n/\mu_s$,
and examples of motions are shown in figure
\ref{fig:TauInfFigure} with $\mu_n/\mu_s$ very small (= $2^{-7}$, left panel), unity (center panel), and very large (=  $2^{7}$, right panel). In each case the body oscillates left and right periodically in time with $\phi$ varying between 0 and $\pi/2$. It can be shown that the body trajectories are independent of how time is parametrized in this limit \cite{AvRa2008a,HaCh2011a,HaBuHoCh2011a,JiAl2013,bittner2018geometrically,alben2019efficient}, so figure \ref{fig:TauInfFigure} also indicates the motions when $\phi$ varies periodically between any
values in the range $[0, \pi/2]$. The motions with $\phi$ ranging from $\pi/2$ to $\pi$ are the same but reflected about the
$y$ axis. Hence the examples of figure \ref{fig:TauInfFigure} are representative of essentially all possible motions when $T \to \infty$. The motions are
dominated by normal motions (left) and tangential motions (right) when the corresponding friction coefficients are 
relatively small. There is an asymmetry between the two limits of frictional anisotropy: the tangential motion can be made arbitrarily small (similar to the left panel), but the normal motion cannot be reduced below a certain magnitude (essentially
that which occurs in the right panel). 

To show that $U \to 0$ when $T \to \infty$,
we again assume
$\delta$ is negligible (as in \cite{JiAl2013}). Since the left hand
side of (\ref{Newtonu}) is zero in this case, (\ref{Newtonu}) reduces to a nonlinear
algebraic equation for $u(\tau)$ that can be solved independently at each $\tau$ in terms of 
$\phi(\tau)$, $d\phi(\tau)/d\tau$, and the given parameters. Given a solution $u(\tau)$, the integral in (\ref{Newtonu}) remains zero when $u(\tau)$ and $d\phi(\tau)/d\tau$ are multiplied by a common factor, so 
solutions
$u(\tau)$ depend linearly on $d\phi(\tau)/d\tau$: 
\begin{align}
u(\tau) = \frac{d\phi}{d\tau}(\tau) \; h\left(\phi(\tau), \mu_n/\mu_s, (\mu_n\mu_s)^{1/4}T\right).
\end{align}
\nn where $h$ is a nonlinear function. The average speed $U$ is proportional to 
\begin{align}
\int_\tau^{\tau+1} u(\tau') d\tau' = \int_\tau^{\tau+1}\frac{d\phi}{d\tau'}(\tau')h\left(\phi(\tau'), \mu_n/\mu_s, (\mu_n\mu_s)^{1/4}T\right) d\tau' =
\int_{\phi(\tau)}^{\phi(\tau+1)} h\left(\phi', \mu_n/\mu_s, (\mu_n\mu_s)^{1/4}T\right) d\phi'
\end{align}
\nn which is zero because $\phi(\tau)$ has period 1. For a nontrivial motion, 
i.e.~one with $d\phi/d\tau \neq 0$ for some $\tau$, $P > 0$, so $\lambda = 0$.

In the opposite limit, $T \to 0$ (fast motions), the motion converges to one with $u(\tau)$ constant
in time, with the constant chosen so that the time-averaged force on the body is zero (which holds for a periodic motion with any $T$ in fact). For a given kinematics, as $T \to 0$, $U$ and $P$ both scale as $1/T$, so
$\lambda$ tends to a constant (possibly 0) in this limit.

\begin{figure} [h]
           \begin{center}
           \begin{tabular}{c}
               \includegraphics[width=5in]{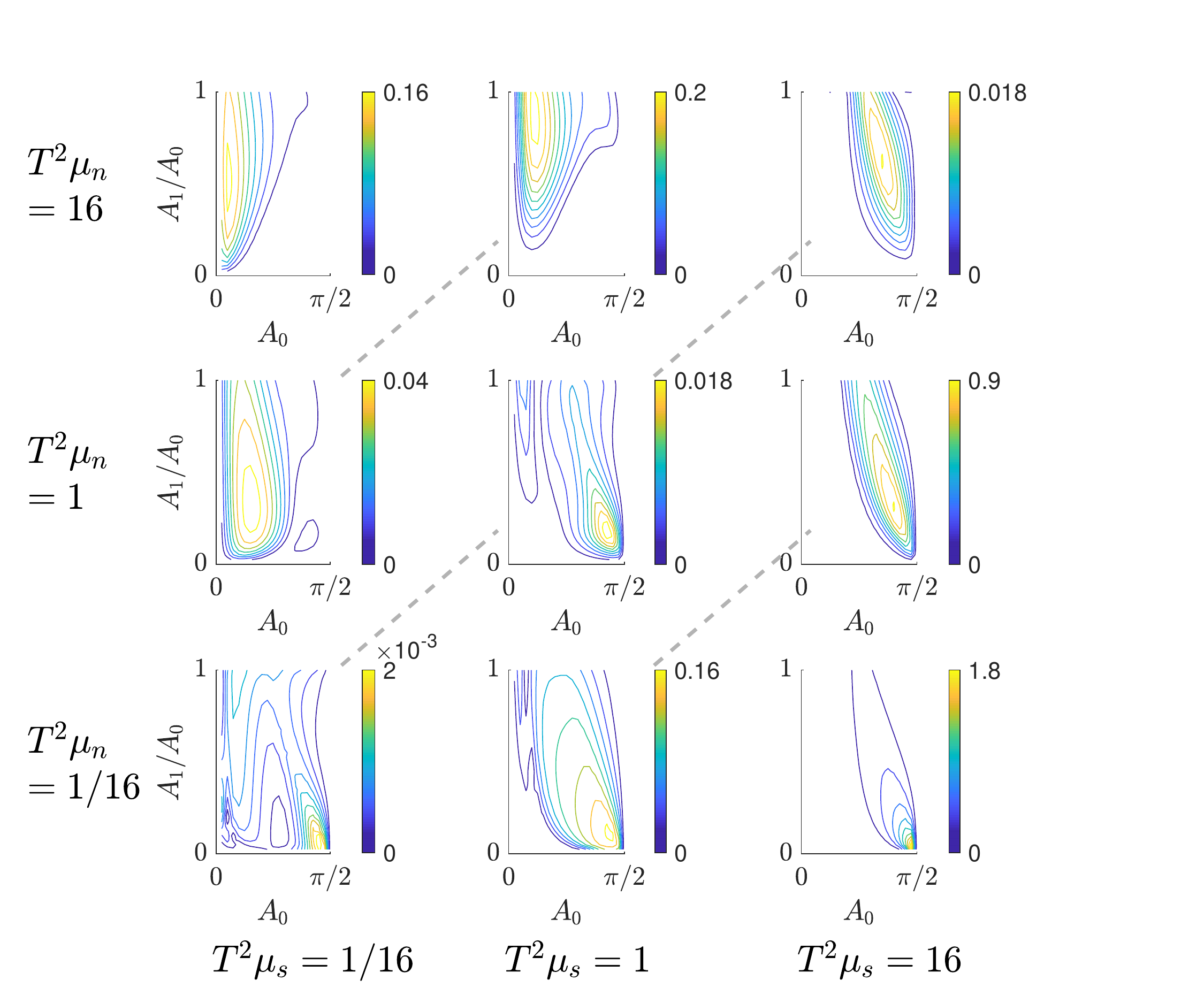} \\
           \vspace{-.25in}
           \end{tabular}
          \caption{\footnotesize
 Contour plots of the magnitude of the scaled efficiency, $\sqrt{\mu_n \mu_s}|\lambda|$, in the four-parameter space of harmonic motions. The contour plots are arranged on a 3-by-3 grid, corresponding to small (1/16), moderate (1), and large (16) values of $T^2\mu_n$ and $T^2\mu_s$ (labeled at left and bottom respectively). At a given pair of these parameters, each contour plot shows the scaled efficiency across values of mean angle $A_0$ and normalized amplitude
$A_1/A_0$. \label{fig:EtaA0A1}}
           \end{center}
         \vspace{-.10in}
        \end{figure}

\section{Harmonic motions \label{sec:Harmonic}}

To gain initial intuition about the motions with finite $T$, 
we first study the simple case where $\phi(\tau)$ is time-harmonic:
\begin{align}
\phi(\tau) = A_0 + A_1 \cos(2\pi \tau),
\end{align}
\noindent with $A_0$ the average angle and $A_1$ the angle oscillation amplitude. The set of
motions with $0 \leq \phi(\tau) \leq \pi$ is described by $0 \leq A_0 \leq \pi/2$ and $0 \leq A_1 \leq A_0$
(those with $\pi/2 \leq A_0 \leq \pi$ are obtained by reflection in the $y$-axis).

Solutions depend on five parameters: $A_0$, $A_1$, $T$, $\mu_n$, and $\mu_s$. In the previous section
we showed that solutions depend on $T$, $\mu_n$, and $\mu_s$ only through the two combinations 
$T^2\mu_s$ and $T^2\mu_n$. In figure \ref{fig:EtaA0A1} we plot contours of the magnitude of the scaled efficiency, $\sqrt{\mu_n \mu_s}|\lambda|$, on a 3-by-3 grid of values of $T^2\mu_n$ and $T^2\mu_s$, labeled at left and bottom respectively. In this limited parameter space, we can plot values for essentially all solutions, and these contour plots give a good representation of the general trends. In each plot, the contours are relatively smooth, and there is clearly a global maximum much higher than any other local maxima (if any exist). The dashed lines link plots with the same $\mu_n/\mu_s$ value. In the upper left region, $\mu_s < \mu_n$, and the optimal $A_0$ is closer to 0 than to $\pi/2$. In the lower right region, $\mu_n < \mu_s$, and the optimal $A_0$ is (much) closer to $\pi/2$ than to 0. Therefore, in each regime the body oscillates near the lowest-drag state ($\phi = 0$ for $\mu_s < \mu_n$ and $\phi = \pi/2$ for $\mu_n < \mu_s$). Along the diagonal from lower left to upper right, drag is isotropic and the contour plots are more complicated. Among all 9 plots, the largest
scaled efficiency, 1.8, occurs in the lower right corner ($\mu_n < \mu_s$), for a motion with small amplitude and $\phi$ close to $\pi/2$. In the upper left region, the largest scaled efficiency is smaller (0.2), and occurs at a motion 
with $\phi$ oscillating between 0 and $\pi/4$. The distinct behaviors for $\mu_n/\mu_s$ large and small indicate that
$\mu_n/\mu_s$ (constant along diagonal dashed lines) is a key parameter. A convenient parameter that varies along each diagonal is the scaled period, $(\mu_n\mu_s)^{1/4}T$. Going forward, we plot results in terms of these two parameters. 

\begin{figure} [h]
           \begin{center}
           \begin{tabular}{c}
               \includegraphics[width=7in]{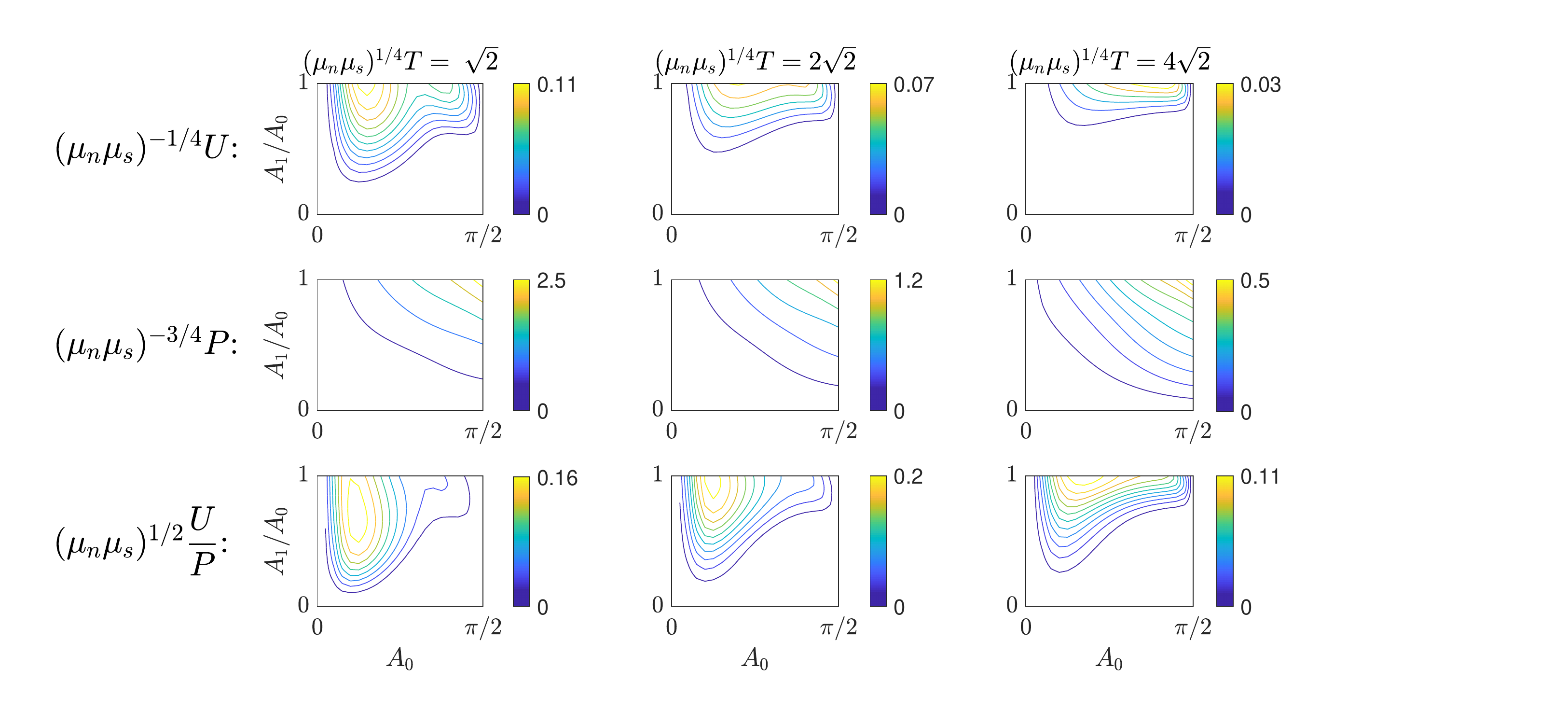} \\
           \vspace{-.25in}
           \end{tabular}
          \caption{\footnotesize Typical patterns of average velocity, input power, and efficiency when normal friction dominates tangential
friction ($\mu_n/\mu_s = 16$ here).
From top to bottom, the rows show contour plots of scaled average velocity $(\mu_n \mu_s)^{-1/4}U$, 
scaled average input power $(\mu_n \mu_s)^{-3/4}P$, and scaled
efficiency $(\mu_n \mu_s)^{1/2}U/P$, respectively. Within each row, results are shown at three values of the scaled oscillation period $(\mu_n \mu_s)^{1/4}T$ (labeled at the top, increasing from left to right). 
 \label{fig:UPEtaDiag1}}
           \end{center}
         \vspace{-.10in}
        \end{figure}

The efficiency plotted in figure \ref{fig:EtaA0A1} is the ratio of average velocity to average power. We now show the behavior of velocity and power separately for $\mu_n/\mu_s$ large and small, by considering two of the diagonals shown in figure \ref{fig:EtaA0A1}: $\mu_n/\mu_s = 2^4$ (upper left dashed line) and $2^{-4}$ (lower right dashed line).
At $\mu_n/\mu_s = 2^4$, figure \ref{fig:UPEtaDiag1} shows the scaled average velocity (top row), power (middle row), and efficiency (bottom row), at three values of the scaled period $(\mu_n\mu_s)^{1/4}T$ (left to right).  The maximum
velocity occurs in the left column of the top row, with $A_0$ near $\pi/6$. When the period is increased (moving rightward), the $A_0$ that maximizes $U$ increases to nearly $\pi/2$, with the maximizing $A_1/A_0$ always the largest value (1). The middle row shows that the power increases monotonically with $A_0$ and $A_1/A_0$ in each case, when the other parameter is held fixed. Trading off between these effects, in the bottom row the maximum scaled efficiency occurs at $A_0$ that remains nearly constant, near $\pi/6$ and 
$A_1/A_0$ values slightly below those that maximize the velocity. 

\begin{figure} [h]
           \begin{center}
           \begin{tabular}{c}
               \includegraphics[width=8in]{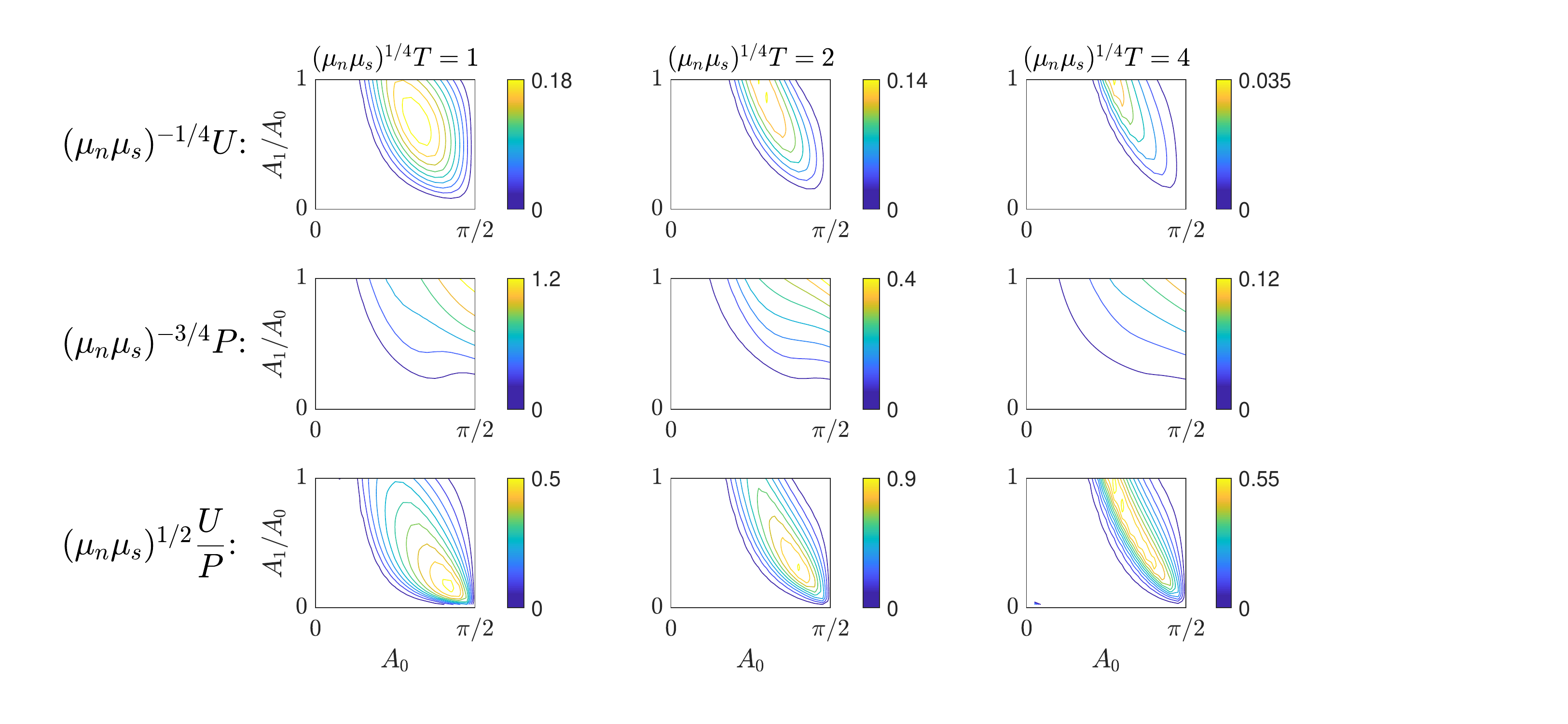} \\
           \vspace{-.25in}
           \end{tabular}
          \caption{\footnotesize Typical patterns of average velocity, input power, and efficiency when tangential friction dominates normal
friction ($\mu_n/\mu_s = 1/16$ here).
From top to bottom, the rows show contour plots of scaled average velocity $(\mu_n \mu_s)^{-1/4}U$, 
scaled average input power $(\mu_n \mu_s)^{-3/4}P$, and scaled
efficiency $(\mu_n \mu_s)^{1/2}U/P$, respectively. Within each row, results are shown at three values of the scaled oscillation period $(\mu_n \mu_s)^{1/4}T$ (labeled at the top, increasing from left to right). 
 \label{fig:UPEtaDiag2}}
           \end{center}
         \vspace{-.10in}
        \end{figure}

With the ratio $\mu_n/\mu_s$ inverted (now $2^{-4}$), figure \ref{fig:UPEtaDiag2} shows the same quantities. 
The velocity (top row) is now maximized with $A_0$ somewhat above $\pi/4$ in each case, with $A_1/A_0$ almost at the maximum in each case. The power (middle row) is nearly monotonic with both parameters, except near $A_0$ = $\pi/2$ at smaller $A_1/A_0$, reflecting the decreased drag when the body is vertical ($\phi = \pi/2$), so the body translates normal to itself.  Combining
these effects, the efficiency (bottom row) is maximized for small $A_1/A_0$ and $A_0$ somewhat below $\pi/2$ (i.e.~small amplitude motions that reach the vertical state, $\phi = \pi/2$).  The elongated contours in the bottom
right panel show that at larger periods, a range of such motions, with both large and small amplitudes, can have nearly the same efficiency.

\begin{figure} [h]
           \begin{center}
           \begin{tabular}{c}
               \includegraphics[width=6in]{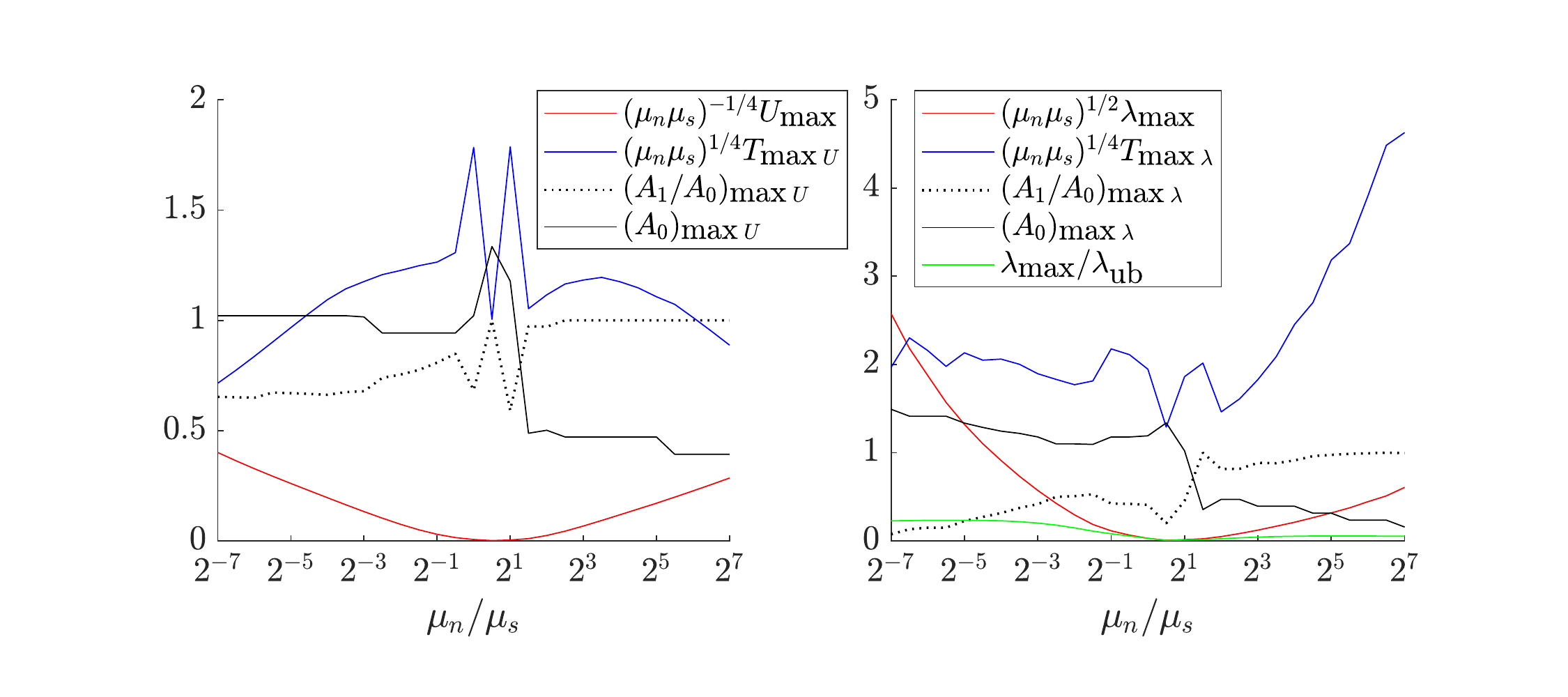} \\
           \vspace{-.25in}
           \end{tabular}
          \caption{\footnotesize Harmonic motion parameters ($A_0$, $A_1/A_0$, $(\mu_n \mu_s)^{1/4}T$) that maximize scaled average speed $(\mu_n \mu_s)^{-1/4}U$ (left) and scaled efficiency $(\mu_n \mu_s)^{1/2}\lambda$ (right),
across a range of friction coefficient ratio ($\mu_n/\mu_s$) values. 
 \label{fig:UEtaMax}}
           \end{center}
         \vspace{-.10in}
        \end{figure}

Figures \ref{fig:UPEtaDiag1} and \ref{fig:UPEtaDiag2} showed velocity and efficiency data at two
$\mu_n/\mu_s$ values, one large and one small. We use similar data across a range of
$\mu_n/\mu_s$, from $2^{-7}$ to $2^{7}$, to determine the maximum scaled
speeds and efficiencies, and the kinematic parameters ($A_0$, $A_1/A_0$, $(\mu_n \mu_s)^{1/4}T$) at which they occur.  Figure \ref{fig:UEtaMax} plots these values for maximum speed (left) and efficiency (right). The maximum speed (red line at left) varies smoothly, with a minimum value of 0.001 slightly away from isotropic friction, at $\mu_n/\mu_s = 1.5$. 
The optimal values of $A_0$ and $A_1/A_0$ are nearly flat for $\mu_n/\mu_s \ll 1$ and $\gg 1$, with
a rapid fluctuation near isotropic friction. The scaled period (blue line) has the same fluctuation, which occurs
because there the velocity is nearly zero for all motions, and there are many small local maximizers.
For each $\mu_n/\mu_s$ value, there is an optimal period, and the velocity tends to zero both as 
$(\mu_n \mu_s)^{1/4}T \to 0$ and $\infty$.

The maximum scaled efficiency (right panel, red line) has a local minimum at $\mu_n/\mu_s = 1.5$, very close to that for velocity. The most efficient motions occur at periods (right panel, blue line) usually about 2 to 4 times those of the fastest motions (left panel, blue line). The kinematic parameters again separate into two different behaviors on either
side of isotropic friction. The maximum relative efficiency, $\lambda/\lambda_{\mbox{ub}}$ (green line), is about 
0.23 for $\mu_n/\mu_s \ll 1$, and 0.05 for $\mu_n/\mu_s \gg 1$. In other words, harmonic motions of a two-link
body can achieve almost one quarter of the maximum possible efficiency for any body with any kinematics, when $\mu_n/\mu_s \ll 1$. For other $\mu_n/\mu_s$ the efficiency is much lower.

\begin{figure} [h]
           \begin{center}
           \begin{tabular}{c}
               \includegraphics[width=6in]{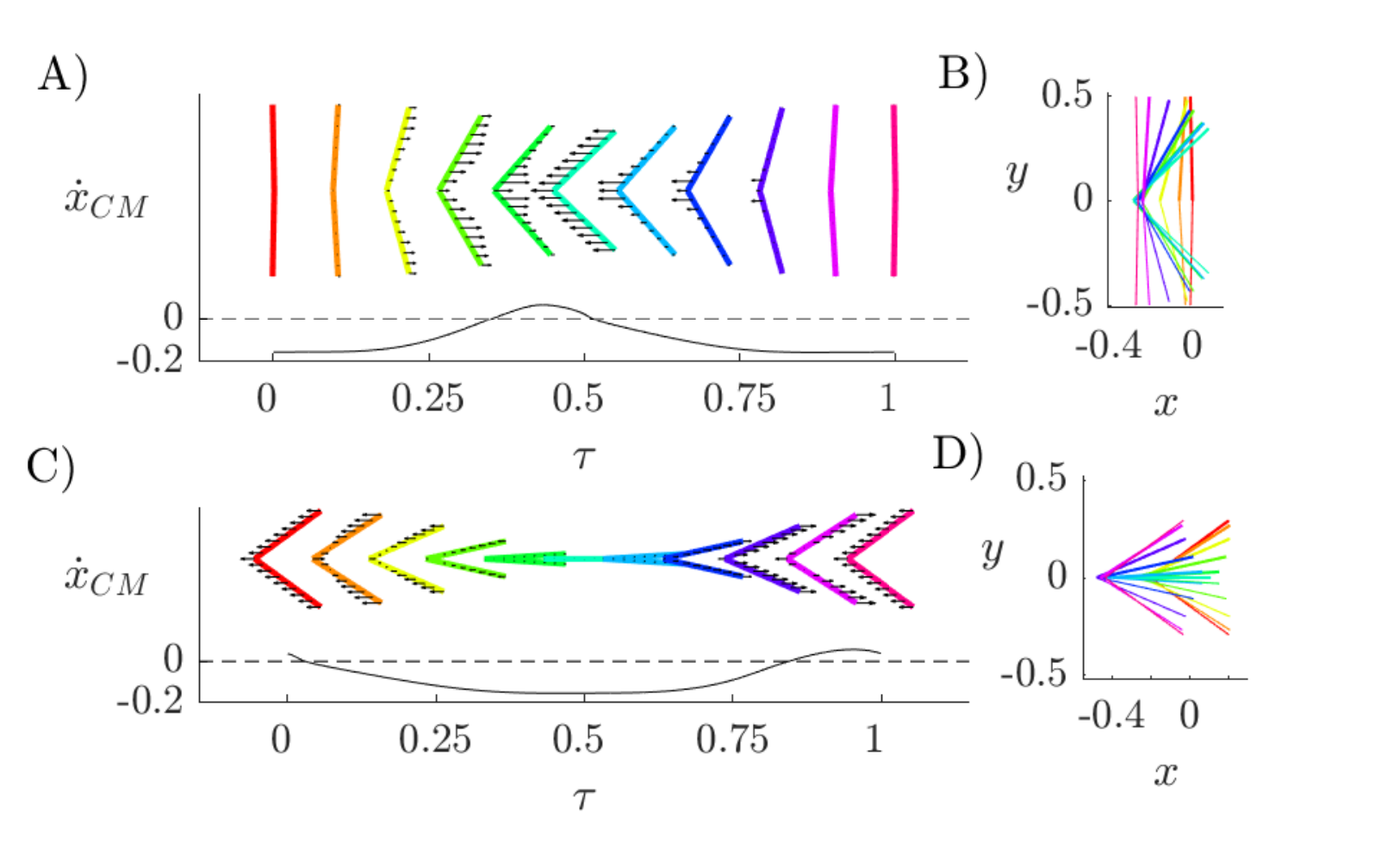} \\
           \vspace{-.25in}
           \end{tabular}
          \caption{\footnotesize Examples of nearly optimally efficient harmonic motions when tangential friction 
is dominant ((A)--(B), $\mu_n = 2^{-8}$, $\mu_s = 2^{0}$, $T = 2^{1.5}$), and when normal friction 
is dominant ((C)--(D), $\mu_n = 2^{0}$, $\mu_s = 2^{-8}$, $T = 2^{1.5}$). In panels A and C, the body velocity
over a period is plotted, with snapshots of the body (colored lines) and horizontal force distributions (small black arrows
extending from points on the body snapshots) at equally spaced time intervals. In panels B and D, the same 
body snapshots are shown in physical space. 
 \label{fig:OptHarmonicMotions}}
           \end{center}
         \vspace{-.10in}
        \end{figure}

We have seen that efficient harmonic motions generally involve the body moving near its lowest drag state ($\phi = 0$ for $\mu_n/\mu_s \gg 1$ and $\phi = \pi/2$ for $\mu_n/\mu_s \ll 1$). In such motions, it is not obvious how the body obtains the thrust force that propels it forward, however. It is also unclear
how the body's translational motion $\dot{x}_{CM}$ evolves during a harmonic oscillation of $\phi$. In figure
\ref{fig:OptHarmonicMotions} we present two optimal motions, one with $\mu_n/\mu_s \ll 1$ (A--B), and
the other with $\mu_n/\mu_s \gg 1$ (C--D). Panel A shows body snapshots (colored lines) with the horizontal force distributions along them (black arrows) at equal intervals over a period, together with
 $\dot{x}_{CM}(\tau)$ (black line), for $\mu_n/\mu_s = 2^{-8}$. Contrary to jet-propelled swimmers, here 
the power stroke occurs when the body opens (blue and purple snapshots), accelerating it leftward. Then the body
``coasts" with little drag in the vertical state (red and orange snapshots), and a nearly constant
$\dot{x}_{CM}(\tau)$. The body then performs the recovery stroke as it closes (yellow and green snapshots). The motion
is shown in physical space in panel B, starting from the red line (far right). As the body closes to the green shape,
it decelerates until it has a slight rightward velocity. As it reopens (green, blue, and purple shapes), most of the body away from the apex moves almost normal to itself, incurring little thrust or drag. But near the apex, the body
moves rightward, resulting in a frictional force that propels it leftward.

In panels C and D, $\mu_n/\mu_s = 2^{8}$, and the situation is largely reversed, and similar to the case with jet-propelled swimmers (for which fluid resistance is greater for normal than tangential motions). On the power
stroke, the body closes (red, orange, yellow, and green snapshots); the body coasts in the low-drag horizontal state (blue-green snapshots), and then has a rightward force as it reopens (purple). Because of the frictional anisotropy,
the body gains and loses equal amounts of momentum on the two strokes even with a net leftward motion.

These results have shown that during the optimal harmonic motions, the body oscillates between the lowest drag state
and another nearby state. We have assumed sinusoidal oscillatory motion between these two states, but now consider other kinematic patterns, to determine whether the body could move more efficiently between these two states.

\section{Power-law motions \label{sec:PowerLaw}}

\begin{figure} [h]
           \begin{center}
           \begin{tabular}{c}
               \includegraphics[width=5in]{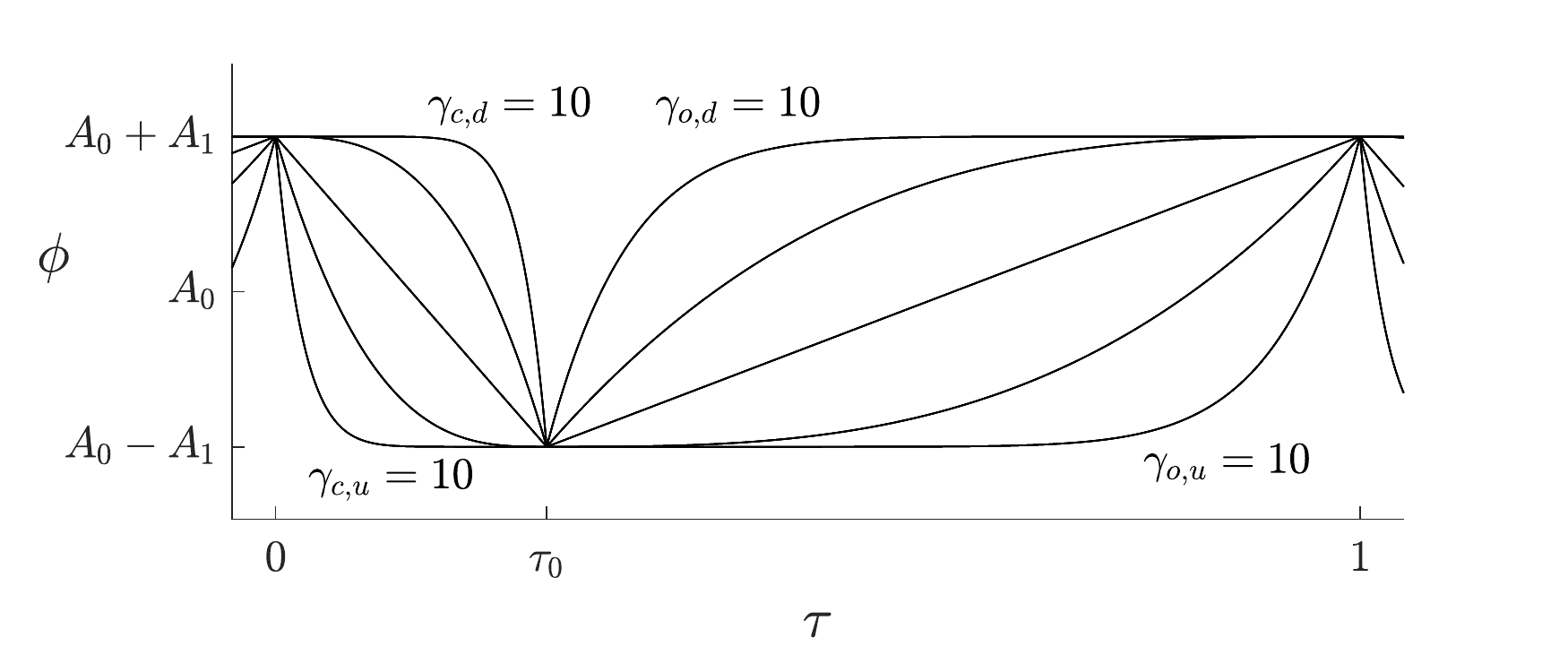} \\
           \vspace{-.25in}
           \end{tabular}
          \caption{\footnotesize Two-phase power law profiles for opening angle functions of time ($\phi(\tau)$).
On the closing phase, from bottom to top, the five curves correspond to the five power laws $\gamma_{c,u} = 10$, 
$\gamma_{c,u} = 3$, $\gamma_{c,u} = \gamma_{c,d} = 1$, $\gamma_{c,d} = 3$, and 
$\gamma_{c,d} = 10$. On the opening phase, the power law values are the same,
with $\gamma_o$ in place of $\gamma_c$.
 \label{fig:TwoPhaseSchematicFigure}}
           \end{center}
         \vspace{-.10in}
        \end{figure}

In order to consider a wider class of kinematics, it is convenient to switch from a sinusoidal motion to one
that is described piecewise. It seems intuitively reasonable to restrict to motions that (like the sinusoidal motions) 
consist of just two phases, opening and closing. Motions with multiple opening and closing phases might 
be expected to experience
additional fluctuations in thrust and drag forces without net improvements in efficiency over motions with just one opening phase and one closing phase.
We will address this hypothesis further using an optimization algorithm in section \ref{sec:Optimal}.

In each phase, we consider a range of kinematics that are power-law functions of time, taken from
\cite{alben2013efficient}, and shown in figure
\ref{fig:TwoPhaseSchematicFigure}. As with the sinusoidal motion, $\phi$ varies between
a maximum and minimum whose average and difference are denoted $A_0$ and $2A_1$, respectively. 
The length of the closing phase
relative to the period is $\tau_0$, sometimes called the ``duty cycle" \cite{moslemi2009effect,peng2012effects,hamlet2012feeding,martin2017flap}. Unlike the sinusoidal motions, the
closing and opening phases may have unequal length (when $\tau_0 \neq 1/2$), allowing different average speeds of
closing and opening. We also allow $\phi$ to remain near its maximum or minimum for longer
times (e.g.~coasting in a low-drag state), by following power laws $\gamma_c$ and $\gamma_o$ 
during the closing and opening phases respectively.

Thus, in the closing phase,
the tangent angle decreases from $A_0 + A_1$ to $A_0 - A_1$ according to
\begin{align}
\phi(\tau) = A_0 + A_1 - 2A_1\left(\frac{\tau}{\tau_0}\right)^{\gamma_{c,d}}, \quad \gamma_{c,d} \geq 1, \quad 0 \leq \tau \leq \tau_0. \label{rapidendc}
\end{align}
\nn Here the exponent $\gamma_{c,d}$ has two subscripts: $c$ for ``closing'' and
$d$ for ``concave {\it down}.'' The corresponding $\phi(\tau)$ plots are those in figure
\ref{fig:TwoPhaseSchematicFigure} which are concave down on $0 \leq \tau \leq \tau_0$.
The highest curve, with $\gamma_{c,d} = 10$, is labeled.
The lower set of curves on $0 \leq \tau \leq \tau_0$ are those which are concave up,
with kinematics given by
\begin{align}
\phi(\tau) = A_0 + A_1 - 2A_1\left(1-\left(1-\frac{\tau}{\tau_0}\right)^{\gamma_{c,u}}\right), \quad \gamma_{c,u} \geq 1, \quad 0 \leq \tau \leq \tau_0.
\label{rapidbegc}
\end{align}
Here the subscripts on $\gamma_{c,u}$ are $c$ for ``closing'' and $u$
for ``concave {\it up}.'' These strokes close most rapidly at the beginning, when $\phi$ is large. The lowest curve in figure \ref{fig:TwoPhaseSchematicFigure} on $0 \leq \tau \leq \tau_0$, with $\gamma_{c,u} = 10$, is labeled.
At the interface between the concave-down and concave-up kinematics is a
straight-line trajectory given both by (\ref{rapidendc}) with
$\gamma_{c,d} = 1$ and (\ref{rapidbegc}) with $\gamma_{c,u} = 1$.

The tangent angles on the opening phase are described analogously, but with the
subscript $o$, for ``opening.'' Those that are concave down on the opening phase increase from $A_0 - A_1$ to $A_0 + A_1$ according to
\begin{align}
\phi(\tau) = A_0 - A_1 + 2A_1\left(1-\left(\frac{1- \tau}{1-\tau_0}\right)^{\gamma_{o,d}}\right), \quad \gamma_{o,d} \geq 1, \quad \tau_0 \leq \tau \leq 1,
\label{rapidbego}
\end{align}
\nn and those that are concave up are given by
\begin{align}
\phi(\tau) = A_0 - A_1 + 2A_1\left(\frac{\tau-\tau_0}{1-\tau_0}\right)^{\gamma_{o,u}},
\quad \gamma_{o,u} \geq 1, \quad \tau_0 \leq \tau \leq 1. \label{rapidendo}
\end{align}
\nn These curves are shown in figure \ref{fig:TwoPhaseSchematicFigure} for $\tau_0 \leq \tau \leq 1$.

\begin{figure} [h]
           \begin{center}
           \begin{tabular}{c}
               \includegraphics[width=7in]{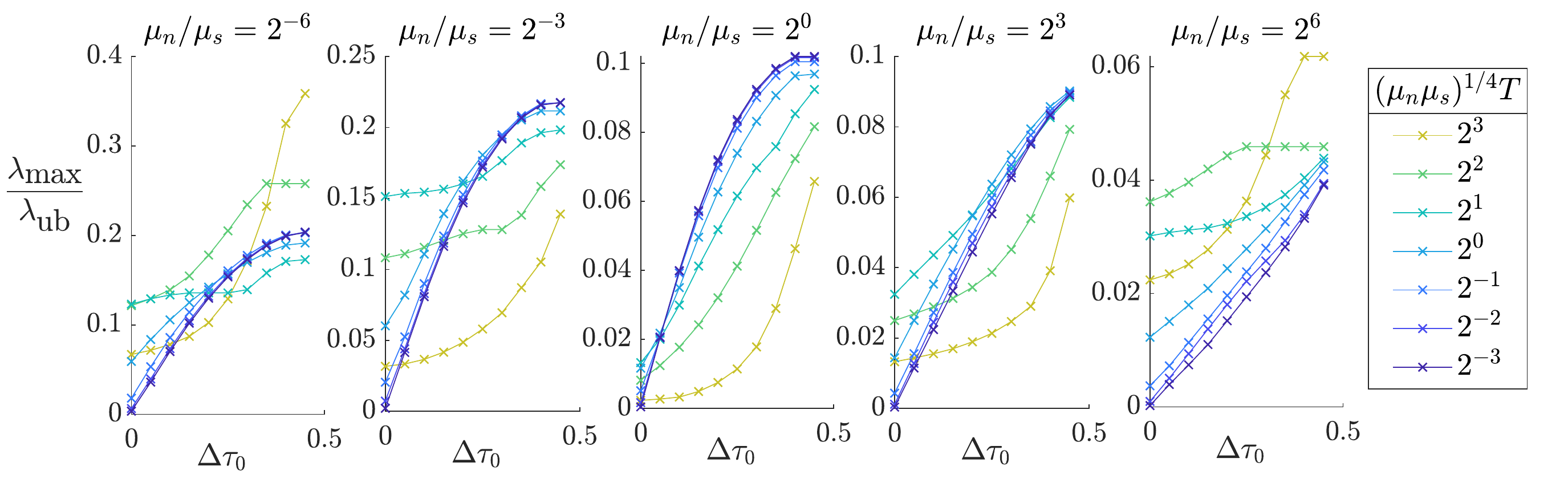} \\
           \vspace{-.25in}
           \end{tabular}
          \caption{\footnotesize Plots of maximum relative efficiency $\lambda_{\mbox{max}}/\lambda_{\mbox{ub}}$
over ranges of the closing phase duration $\tau_0$ of length 2$\Delta\tau_0$. The set of $\tau_0$ includes more asymmetric opening and closing phases as $\Delta\tau_0$ increases. Here $\tau_0$ is allowed to vary over the interval [$0.5 -\Delta\tau_0$, $0.5 +\Delta\tau_0$], and the tangent angle functions are constrained to be linear in time
(i.e.~$\gamma_{c,u}$ are $\gamma_{o,u}$ both fixed at 1). In other words, $\tau_0$ is allowed
to vary over an interval of length $2\Delta\tau_0$ centered at 0.5 (the symmetric case of closing and opening phases with equal duration).  \label{fig:EtaVsMint0}}
           \end{center}
         \vspace{-.10in}
        \end{figure}

We first consider the effect of varying $\tau_0$, together with the two parameters $A_0$ and $A_1$ already considered
for sinusoidal strokes. We keep $\gamma_{c} = \gamma_{o} = 1$, so we are replacing sinusoidal with
linear $\phi$, but in sample cases we do not find large qualitative changes due to this change in functional form.
We start by studying the change in maximum relative efficiency when $\tau_0$ is allowed to range over intervals
[$0.5 -\Delta\tau_0$, $0.5 +\Delta\tau_0$], with $\Delta\tau_0$ increasing from 0 (allowing only the symmetric duty cycle 0.5) to 0.45 (allowing nearly the whole range). In each case $A_0$ and $A_1/A_0$ are allowed to vary over their whole ranges ($(0, \pi/2]$ and $(0, 1)$, respectively). In figure \ref{fig:EtaVsMint0} we plot the maximum relative efficiencies thus obtained, for various combinations of $\mu_n/\mu_s$ (labeled at top) and $(\mu_n \mu_s)^{1/4}T$ (in legend at right). By definition, the maximum relative efficiencies are nondecreasing with increasing $\Delta\tau_0$.
In fact, they increase by a factor $\gg 1$ in most cases, showing that $\tau_0$ has a strong effect on efficiency. 
For symmetric strokes ($\Delta\tau_0 = 0$), the efficiency (and velocity) tend to zero as
$(\mu_n \mu_s)^{1/4}T$ tends to zero. For asymmetric strokes, the same is not true. In fact, for $2^{-3} \leq \mu_n/\mu_s \leq 2^3$, the most efficient strokes are obtained with the shortest scaled periods shown, $2^{-3}$. Further decreases lead to little further change in relative efficiency, so fast oscillations are the most efficient in this regime. At the most anisotropic friction ratios ($2^{-6}$ and $2^6$), longer oscillation periods are preferred, but
the efficiencies still increase substantially with $\Delta\tau_0$.

\begin{figure}[!h]
           \begin{center}
           \begin{tabular}{c}
               \includegraphics[width=5.5in]{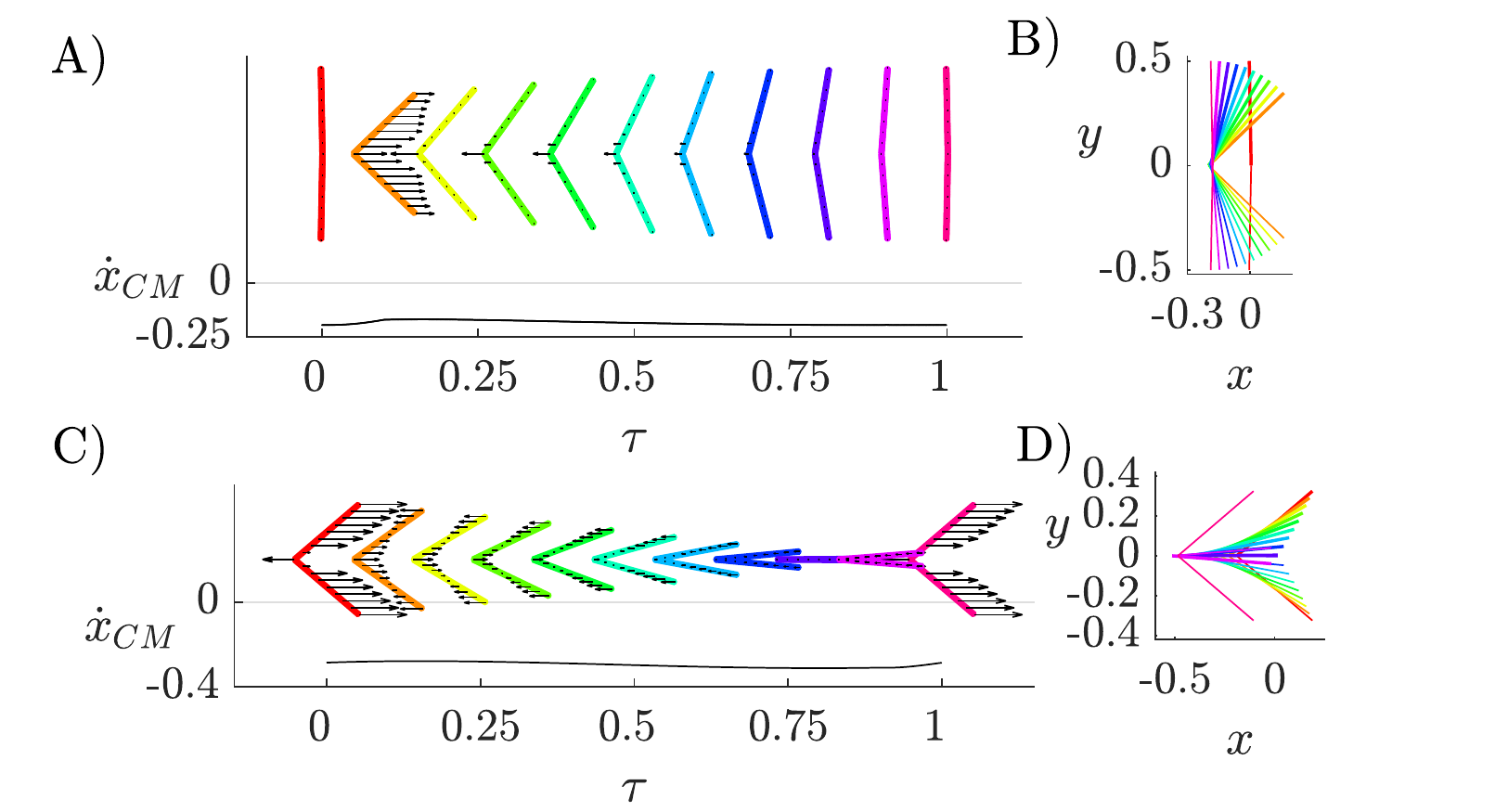} \\
           \vspace{-.25in}
           \end{tabular}
          \caption{\footnotesize  Examples of efficient motions when the tangent angle is a linear function of time during the opening and closing phases. In (A)--(B), tangential friction 
is dominant ($\mu_n = 2^{-8}$, $\mu_s = 2^{0}$), $T = 2^{0}$ and $\tau_0 = 0.1$. In (C)--(D), normal friction 
is dominant ($\mu_n = 2^{0}$, $\mu_s = 2^{-8}$), $T = 2^{0}$ and $\tau_0 = 0.9$. In panels A and C, the body velocity
over a period is plotted, with snapshots of the body (colored lines) and horizontal force distributions (small black arrows
extending from points on the body snapshots) at equally spaced time intervals. In panels B and D, the same 
body snapshots are shown in physical space.
 \label{fig:Optt0Motions}}
           \end{center}
         \vspace{-.10in}
        \end{figure}

\begin{figure}[!h]
           \begin{center}
           \begin{tabular}{c}
               \includegraphics[width=4.5in]{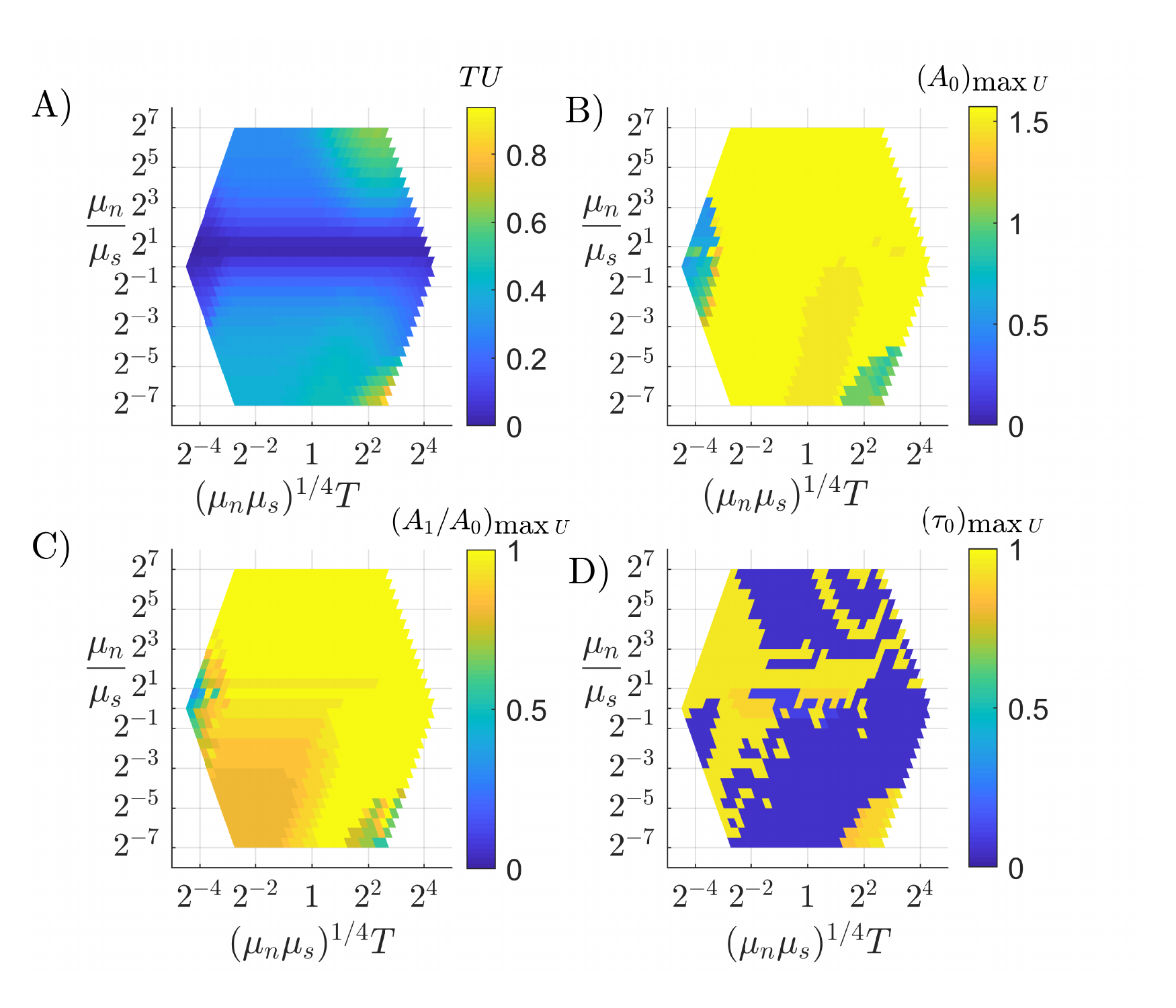} \\
           \vspace{-.25in}
           \end{tabular}
          \caption{\footnotesize Kinematic parameters yielding the fastest motions when the tangent angle is a linear function of time during the opening and closing phases. A) The maximum speed $U$ multiplied
by the oscillation period $T$,
over all $A_0$, $A_1$, and $\tau_0$, for each value of $\mu_n/\mu_s$ and $(\mu_n \mu_s)^{1/4}T$.
The remaining panels show the values of 
(B) $A_0$, (C) $A_1/A_0$, and (D) $\tau_0$ at which the maximum speed is
attained. \label{fig:UOptt0}}
           \end{center}
         \vspace{-.10in}
        \end{figure}

Examples of the most efficient motions over all $\tau_0$ 
(and $A_0$, $A_1$, and $(\mu_n \mu_s)^{1/4}T$) are shown in figure \ref{fig:Optt0Motions},
for $\mu_n/\mu_s = 2^{-8}$ (A--B) and $2^{8}$ (C--D). The motions are essentially the same as those in figure
\ref{fig:OptHarmonicMotions}, but with a much longer power stroke and a much shorter recovery stroke. Because
of the Coulomb force law, even with a very rapid recovery stroke, the magnitude of the drag force is bounded.
Because they act over a short time, the drag forces do not decelerate the body substantially.  
Consequently, the body can reach a larger net velocity at steady state, at which both thrust and drag forces
act on the body during the power stroke (shown by the arrows on the orange, yellow, and green bodies in panel C). The net thrust force is small, but acts over a longer time, and is therefore sufficient to maintain steady 
locomotion at this higher speed.

The kinematic parameters that maximize the average speed over wide ranges of $\mu_n/\mu_s$ and $(\mu_n \mu_s)^{1/4}T$ are plotted in figure \ref{fig:UOptt0}. Panel A plots the maximum values of $TU$ over
$A_0$, $A_1$, and $\tau_0$. The values
remain roughly constant as $(\mu_n \mu_s)^{1/4}T$ becomes small, showing that $U$ scales as $1/T$ in 
this limit, which is typical for asymmetric strokes ($\tau_0 \neq 0.5$). The $A_0$ and $A_1/A_0$ that maximize
$U$ are shown in panels B and C respectively. Each is near its maximum value for most $\mu_n/\mu_s$ and $(\mu_n \mu_s)^{1/4}T$ values. With $A_0$ near $\pi/2$, the strokes move left and right almost symmetrically, but an asymmetric duty cycle allows high speed locomotion in this case. For two special cases, higher speeds are obtained at smaller amplitudes: $\mu_n/\mu_s \ll 1$ and long periods, where the motions are similar to figure \ref{fig:Optt0Motions}B, and a small region near
isotropic friction, with very short oscillation periods. Panel D shows that the highest speeds are obtained when 
the duty cycle is asymmetric in either direction (i.e.~$\tau_0$ close to 0 or to 1), both for $\mu_n/\mu_s <1$
and $> 1$.

\begin{figure}[!ht]
           \begin{center}
           \begin{tabular}{c}
               \includegraphics[width=5in]{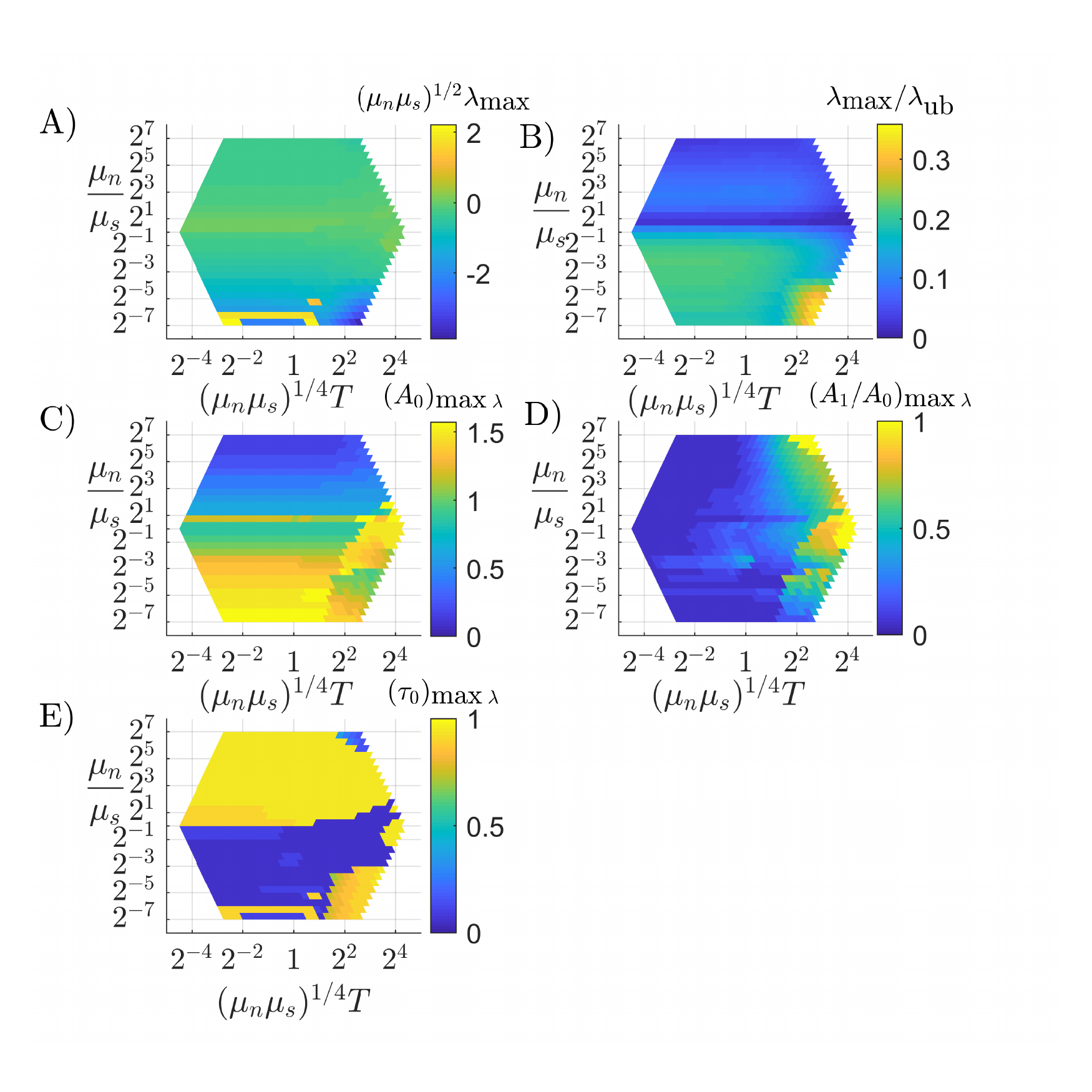} \\
           \vspace{-.25in}
           \end{tabular}
          \caption{\footnotesize Kinematic parameters yielding the most efficient motions when the tangent angle is a linear function of time during the opening and closing phases. A) The maximum scaled efficiency $(\mu_n \mu_s)^{1/2}\lambda$
over all $A_0$, $A_1$, and $\tau_0$, for each value of $\mu_n/\mu_s$ and $(\mu_n \mu_s)^{1/4}T$.
B) The maximum normalized efficiency $\lambda_{\mbox{max}}/\lambda_{\mbox{ub}}$.
The remaining panels show the values of 
(C) $A_0$, (D) $A_1/A_0$, and (E) $\tau_0$ at which the maximum efficiency is
attained.
 \label{fig:EtaOptt0}}
           \end{center}
         \vspace{-.10in}
        \end{figure}

\begin{figure} [!h]
           \begin{center}
           \begin{tabular}{c}
               \includegraphics[width=6.5in]{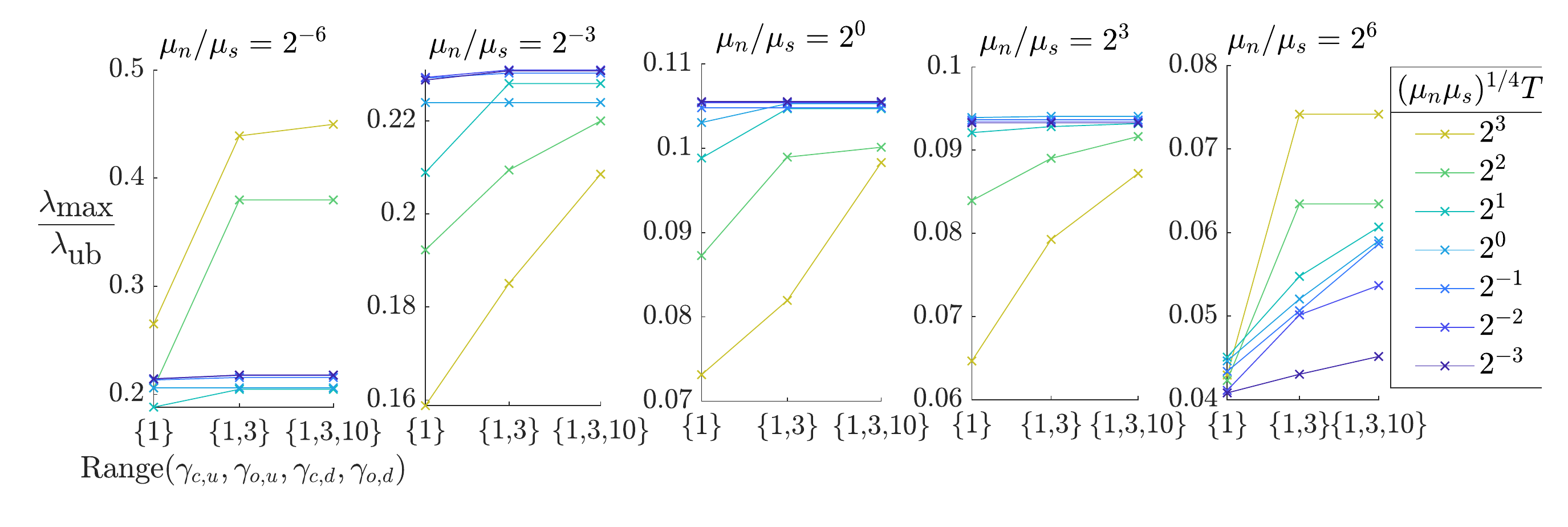} \\
           \vspace{-.25in}
           \end{tabular}
          \caption{\footnotesize Plots of maximum relative efficiency $\lambda_{\mbox{max}}/\lambda_{\mbox{ub}}$
when the tangent angle functions of time are allowed to assume power laws in the sets $\{1\}$, $\{1,3\}$, and
$\{1,3, 10\}$.  In each case $\tau_0$, $A_0$, and $A_1$ are allowed to vary over their full ranges.
 \label{fig:EtaVsMinGamma}}
           \end{center}
         \vspace{-.10in}
        \end{figure}

With large oscillations in the forward and backward directions, the fastest motions are, not surprisingly,
not the most efficient in general. Figure \ref{fig:EtaOptt0} shows the optimal values of
efficiency and the kinematic parameters at which they are attained. Panel A plots the signed efficiency
(i.e.~the sign of $\lambda$ is the sign of $U$). Because $A_0$ is restricted to $[0, \pi/2]$, there is
a bias towards motions with negative $U$ (mirror image motions with positive $U$ would be obtained if we
considered $A_0 \in [\pi/2, \pi]$). In two special cases, however, the most efficient motions move rightward: $\mu_n/\mu_s$ near 1, and near $2^{-6}$, for a wide range of $(\mu_n \mu_s)^{1/4}T$
in both cases. Panel B shows the relative efficiencies (in absolute value). For $\mu_n/\mu_s > 1$ (top half of the hexagonal region), values range from 0.03 to 0.09, with peak values near   
$(\mu_n \mu_s)^{1/4}T = 1$ in most cases. For $\mu_n/\mu_s < 1$ (bottom half), the values
are mostly near 0.2, except in the lower right corner where they reach 0.35, at the largest
$(\mu_n \mu_s)^{1/4}T$ shown. Although the lower right corner gives the highest relative efficiency,
the motions in the lower left corner have more than half this value, and are much faster (since speed scales as $1/T$). Panel C shows that the peak efficiencies occur at $A_0$ that,
as before, are closer to 0 for $\mu_n/\mu_s > 1$ and to $\pi/2$ for $\mu_n/\mu_s < 1$.
The corresponding amplitude values ($A_1/A_0$, panel D) are small for fast oscillations (small $(\mu_n \mu_s)^{1/4}T$),
and larger for slow oscillations (large $(\mu_n \mu_s)^{1/4}T$). Finally, the optimal duty cycle $\tau_0$ (panel E) is mostly near 1 for $\mu_n/\mu_s > 1$ and near 0 for $\mu_n/\mu_s < 1$, like the motions
shown in figure \ref{fig:Optt0Motions}.

We have seen in figure \ref{fig:EtaVsMint0} that varying the duty cycle alone is sufficient to increase efficiencies by a factor of 2 or more in most cases. We now examine the effects of allowing the power laws
during opening and closing, $\gamma_o$ and $\gamma_c$, to vary from 1. In figure \ref{fig:EtaVsMinGamma}, we show the maximum relative efficiencies as 
$\gamma_o$ and $\gamma_c$ are allowed to range over progressively larger sets of values
(similarly to figure \ref{fig:EtaVsMint0} for $\tau_0$), consisting of the powers $\{1\}$, $\{1,3\}$, and
$\{1,3, 10\}$. With small oscillation periods (dark blue lines), there is little improvement beyond
the linear case. The longer the oscillation period, the greater the improvement from considering
more concave or convex functions. Generally, most of the improvement occurs just by adding cubic
functions. Overall, allowing $\gamma_o$ and $\gamma_c$ to change from 1 allows further improvements in peak relative efficiency by factors of 2 ($\mu_n/\mu_s = 2^{-6}$), 1.5 ($\mu_n/\mu_s = 2^{6}$), or
essentially no improvement (the other $\mu_n/\mu_s$ shown).

\begin{figure} [h]
           \begin{center}
           \begin{tabular}{c}
               \includegraphics[width=5.5in]{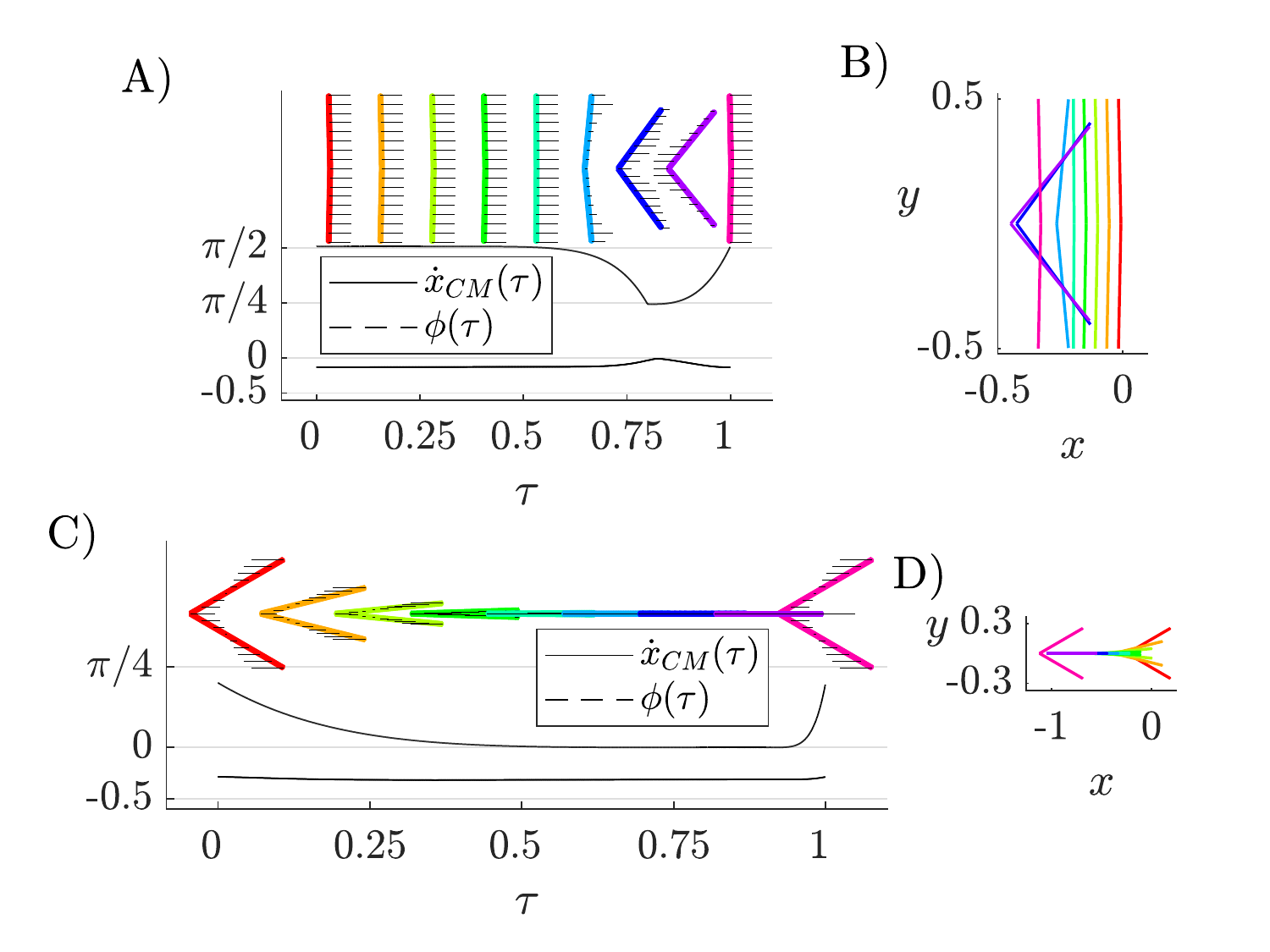} \\
           \vspace{-.25in}
           \end{tabular}
          \caption{\footnotesize Examples of efficient motions when the tangent angle is a power law function of time during the opening and closing phases. In (A)--(B), tangential friction 
is dominant ($\mu_n = 2^{-8}$, $\mu_s = 2^{0}$), $T = 2^{1.5}$, $\tau_0$ = 0.8, $\gamma_{c,d} = 10$, and
$\gamma_{o,u} = 3$. In (C)--(D), normal friction 
is dominant ($\mu_n = 2^{0}$, $\mu_s = 2^{-8}$), $T = 2^{1.5}$), $\tau_0$ = 0.2, $\gamma_{c,u} = 5$, and
$\gamma_{o,u} = 5$. In panels A and C, the tangent angle $\phi(\tau)$ and body velocity $\dot{x}_{CM}(\tau)$
are plotted, with snapshots of the body (colored lines) and horizontal force distributions (small black arrows
extending from points on the body snapshots) at equally spaced time intervals. In panels B and D, the same 
body snapshots are shown in physical space.
 \label{fig:OptAllParamsMotions}}
           \end{center}
         \vspace{-.10in}
        \end{figure}

For the more anisotropic friction regimes, where the power laws yield large improvements, we 
present examples of efficient motions in figure \ref{fig:OptAllParamsMotions}. The motions
are similar to those in figure \ref{fig:Optt0Motions}, except that the concavity/convexity of
$\phi$ (plotted with dashed lines) allows the bodies to ``coast" in their lowest drag states
($\phi = \pi/2$ in panels A and B, and $\phi = 0$ in panels C and D). In panel A, there is a
rapid closing and opening for $0.5 < \tau < 1$. The efficiency is sensitive to the 
detailed form of these rapid motions. For example, changing the opening stroke in panels A--B 
from concave up to concave down can reduce the efficiency by more than a factor of 3.

\begin{figure} [h]
           \begin{center}
           \begin{tabular}{c}
               \includegraphics[width=5in]{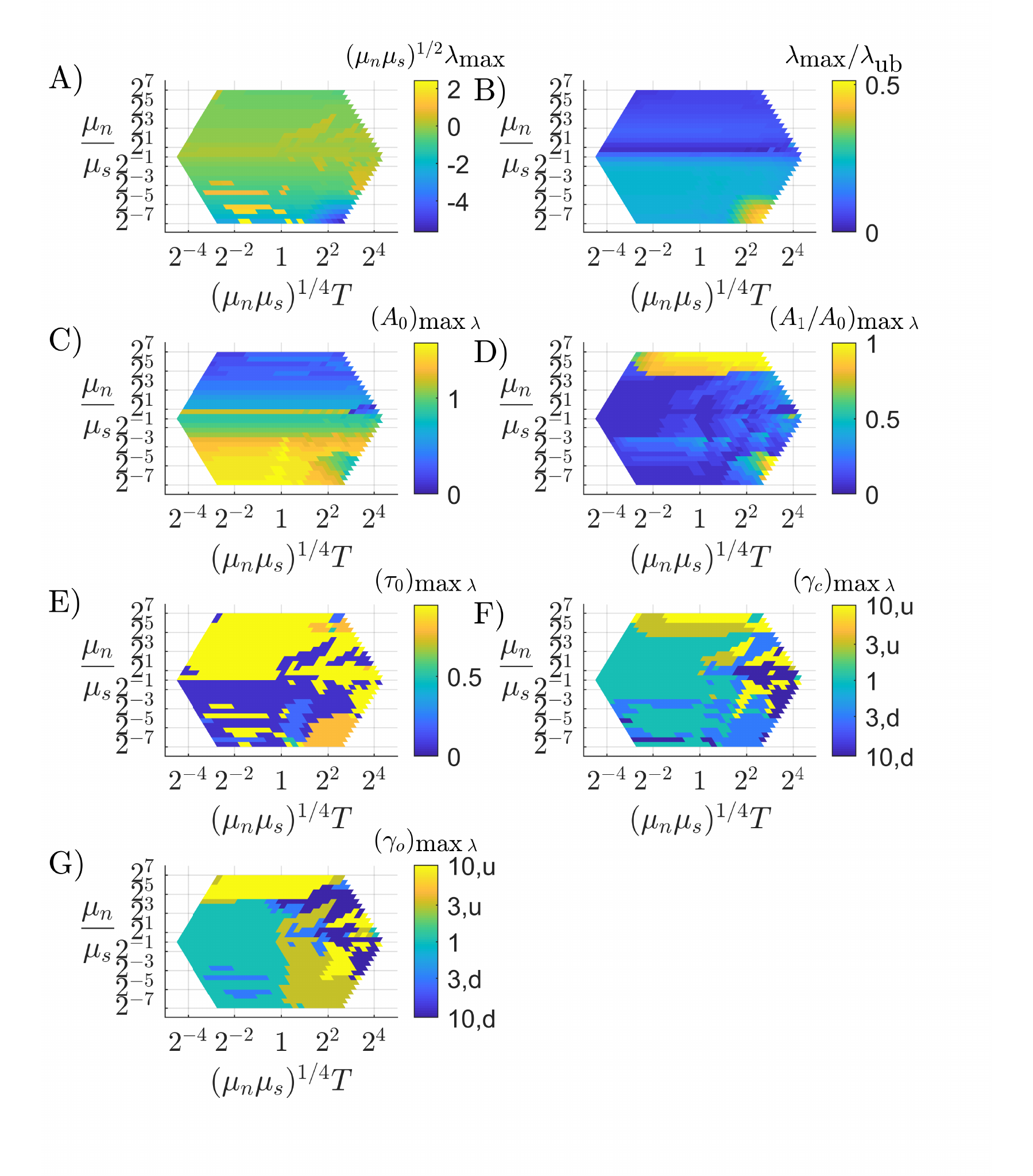} \\
           \vspace{-.25in}
           \end{tabular}
          \caption{\footnotesize Kinematic parameters yielding the most efficient motions when the tangent angle is a power-law function of time during the opening and closing phases. A) The maximum scaled efficiency $(\mu_n \mu_s)^{1/2}\lambda$
over all $A_0$, $A_1$, $\tau_0$, $\gamma_{c}$, and $\gamma_{o}$ for each value of $\mu_n/\mu_s$ and $(\mu_n \mu_s)^{1/4}T$.
B) The maximum normalized efficiency $\lambda_{\mbox{max}}/\lambda_{\mbox{ub}}$.
The remaining panels show the values of 
(C) $A_0$, (D) $A_1/A_0$, (E) $\tau_0$, (F) $\gamma_c$, and (G) $\gamma_o$ at which the maximum efficiency is
attained.
 \label{fig:EtaOptAllParams}}
           \end{center}
         \vspace{-.10in}
        \end{figure}

We present the kinematic parameters that are optimal for efficiency across $\mu_n/\mu_s$ and $(\mu_n \mu_s)^{1/4}T$ space in figure \ref{fig:EtaOptAllParams}. In the more isotropic portion
of parameter space ($2^{-4} < \mu_n/\mu_s < 2^4$), the values of scaled
efficiency (A) and relative efficiency (B) are only about 0--10\% higher than the corresponding
values with $\phi$ constrained to be linear (figure \ref{fig:EtaOptt0}), also indicated by
figure \ref{fig:EtaVsMinGamma}. For $1 < \mu_n/\mu_s < 2^4$, figure \ref{fig:EtaOptAllParams}D
shows that the optimal amplitude ratio $A_1/A_0$ is much less than 1, so here the optimal motions
do not coast in the minimal drag state ($\phi =0$, which is only reached when $A_1/A_0 = 1$), but
instead undergo small, rapid oscillations about mean $\phi$ in the range $\pi/6$--$\pi/4$ (shown by the
corresponding $A_0$ values in figure \ref{fig:EtaOptAllParams}C).
As $\mu_n/\mu_s$ increases above $2^4$, burst-and-coast motions appear with $A_1/A_0 = 1$, and
the peak relative efficiency increases by about a factor of 2 over the linear case, to 0.06. For 
$\mu_n/\mu_s < 1 $ the peak relative
efficiencies increase by about 10\% over the linear case except in the lower right corner,
where the efficiency increases by about 50\%. Comparing the kinematic parameters at which
these efficiencies occur with those in figure \ref{fig:EtaOptt0}, we find little change in $A_0$ (panel C), and somewhat more change in $A_1/A_0$ (panel D), with larger relative
amplitudes for $\mu_n/\mu_s \gg 1$. The duty cycle (panel E) still shows a basic divide
between small values at small $\mu_n/\mu_s$ and large values at large $\mu_n/\mu_s$, but
with more exceptions now. The optimal opening and closing power laws (panels F and G) 
are 1 in a large portion of parameter space where $(\mu_n \mu_s)^{1/4}T < 1$, with 
increasingly concave/convex values (increasing from 3 to 10) as $(\mu_n \mu_s)^{1/4}T$
becomes larger. For $(\mu_n \mu_s)^{1/4}T > 1$, the variations in $\tau_0$, $\gamma_o$, and $\gamma_c$ are somewhat correlated with each other. In some cases these correspond to changes from leftward
 to rightward motions (shown by changes from green to orange in panel A). In other cases, they are different
versions of burst-and-coast motions, which can be achieved both with $\tau_0$ close to 0 and close to 1.
In either case, choosing $\gamma_o$ (or $\gamma_c$) concave up or concave down allows the body to coast 
with $\phi = 0$ or $\pi/2$, respectively. 




\section{Optimal motions \label{sec:Optimal}}

Thus far we have studied motions that consist of two phases (opening and closing), described by harmonic or power law functions of time. The functions are described by up to 5 parameters, and we are able to compute motions covering essentially
the full ranges of values of all the parameters. We now study optimal motions in a larger and more general parameter space, that can approximate essentially any smooth periodic function (when sufficiently many parameters are used). These optima may indicate whether there are optimal motions completely different from those we have found so far by assuming a two-phase motion, and to what extent further gains in efficiency are possible. With more parameters, it is no longer feasible to compute motions throughout parameter space, so we use a local optimization search, but starting from a large number of random initial kinematic parameters in order to identify the global optima. 

Specifically, we employ a quasi-Newton optimization algorithm (BFGS, with frequent
Newton steps) to determine $\phi(\tau)$ that maximize $\lambda$ in spaces of truncated Fourier series:
\begin{align}
\phi(\tau) = A_0 + \sum_{k = 1}^{K_{\mbox{max}}} A_k \cos(2\pi k \tau) + B_k \sin(2\pi k \tau).  \label{Fourier}
\end{align}
\nn In (\ref{Fourier}) there are $2K_{\mbox{max}} + 1$ parameters; adding the oscillation period $T$ (which appears in (\ref{Newtonu})) gives $2K_{\mbox{max}} + 2$ parameters to maximize over. One can reduce the number of parameters by
one by setting $B_1$ to zero (which sets an arbitrary phase of the motion). We use the semi-implicit linearized iteration mentioned in section \ref{sec:Model} (and described in \cite{alben2019efficient}) to compute $\lambda$. We write an exact formula for the gradient of $\lambda$ with respect to the parameters using repeated applications of the chain rule. The gradient is used to compute a low-rank approximate Hessian matrix (for BFGS steps) or a finite-difference Hessian matrix (for Newton steps). We use a damped line search to
select updates that decrease $\lambda$, and iterate until the gradient norm is below a preset tolerance (usually $10^{-7}$). By minimizing $\lambda$, the algorithm chooses motions with $U < 0$, but there is no loss of generality here---reflecting the motions about $\phi = \pi/2$ takes $U$ to $-U$.

We run the optimization with a variety of $K_{\mbox{max}}$, showing results for 1, 3, 7, and 10 here. As $K_{\mbox{max}}$ increases, we observe (not surprisingly) that the algorithm converges more slowly and to a wider range of local optima. Unlike the choices of
$\phi$ in previous sections, the coefficients in (\ref{Fourier}) are arbitrary, so $\phi$ can take on values outside the interval
$[0, \pi]$. If $\phi \in ((2n-1)\pi, 2n\pi), n \in \mathbb{Z}$ the body segment with $0 \leq s \leq 1/2$ in figure \ref{fig:ScallopSchematic} 
lies in the lower half plane. Equations (\ref{Newton}) and (\ref{power}) remain valid in this case, because
they assume that the portion with $-1/2 \leq s \leq 0$ is the
reflection in the $x$-axis of that with $0 \leq s \leq 1/2$. Thus if $\phi \in ((2n-1)\pi, 2n\pi)$, we can assume
that $0 \leq s \leq 1/2$ now describes the lower half of the body. When $\phi$ passes through
$n\pi$, instead of the two halves of the body passing through one another, we assume that they contact and reverse
directions, so that the upper half and lower half always remain in their respective half planes. At such instants
of contact, $d\phi/d\tau$ is generally nonzero, and thus the angular velocities of the top and bottom each
jump to the negative of their values before contact. Such kinematics are physically valid,
and do not introduce theoretical or computational obstacles for solving the equations. 
The resulting $u(\tau)$ appear to be smooth in such cases. 

\begin{figure} [h]
           \begin{center}
           \begin{tabular}{c}
               \includegraphics[width=6in]{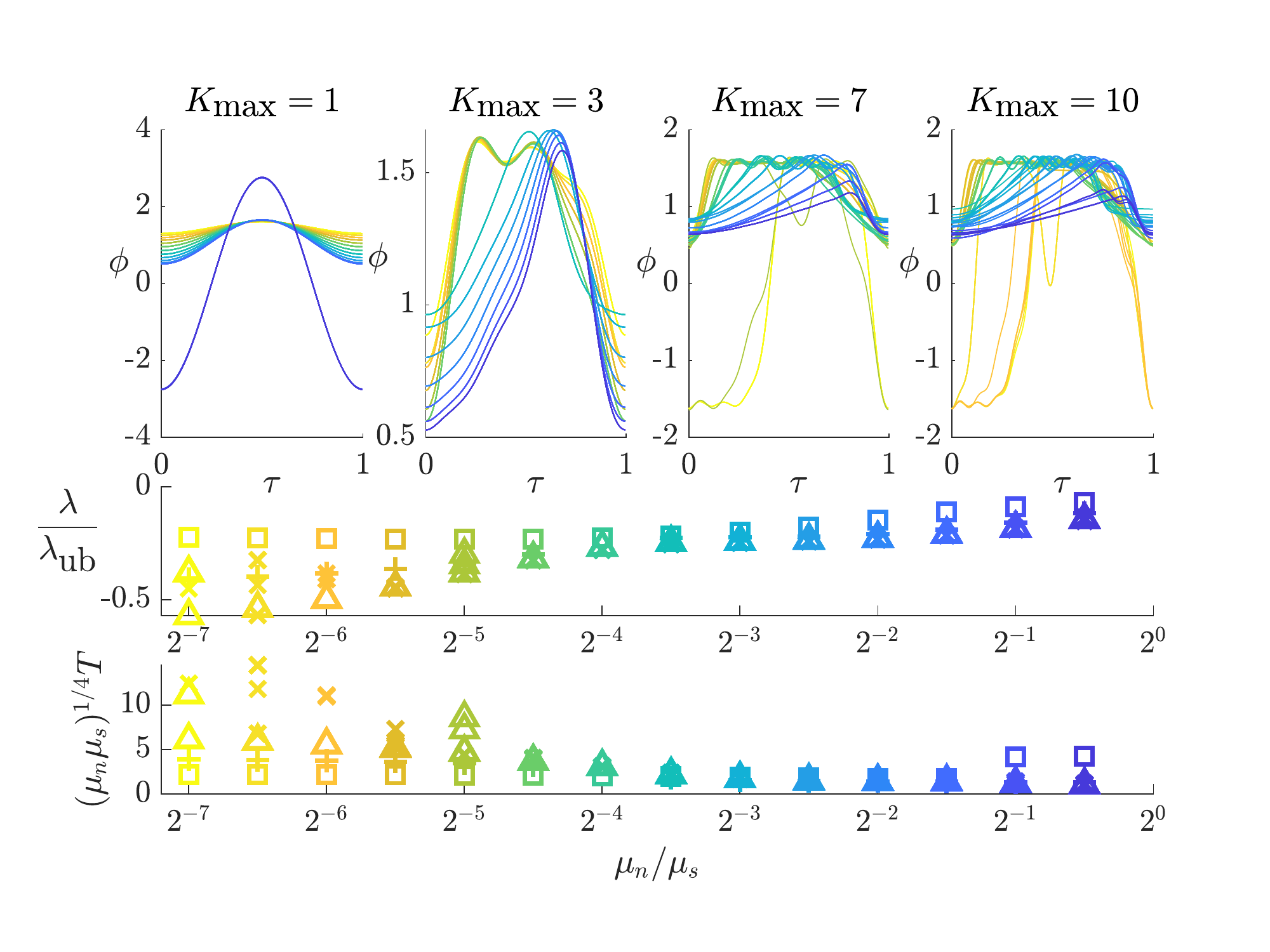} \\
           \vspace{-.25in}
           \end{tabular}
          \caption{\footnotesize Optimal motions in the space of truncated Fourier series when tangential friction is dominant ($\mu_n/\mu_s < 1$). Top row: optimal tangent angle functions $\phi(\tau)$ in the space of Fourier series with wavenumbers up to $K_{\mbox{max}}$ = 1, 3, 7, and 10. Colors correspond to friction ratios $\mu_n/\mu_s$, labeled in the middle and bottom rows. Corresponding normalized efficiencies $\lambda/\lambda_{\mbox{ub}}$ (middle row)
and scaled periods (bottom row) are plotted versus $\mu_n/\mu_s$ for $K_{\mbox{max}}$ = 1 (squares), 3 (plusses), 7 (triangles), and 10 (crosses).
 \label{fig:OptMotionsLowMun}}
           \end{center}
         \vspace{-.10in}
        \end{figure}

We initialize a large ensemble of $\phi$ in (\ref{Fourier}) with $A_0$ and $A_1$ randomly chosen on the unit interval, and the remaining coefficients zero. If instead all coefficients are set to random values, the algorithm spends more time near suboptimal motions dominated by high frequencies, and thus converges less often and more slowly with large $K_{\mbox{max}}$. In figure \ref{fig:OptMotionsLowMun} we plot the ten most efficient local optima (with repetition) for various 
choices of $\mu_n/\mu_s < 1$, and $K_{\mbox{max}}$ = 1, 3, 7, and 10 (top row). The colors denote the
values of $\mu_n/\mu_s$, labeled by the horizontal axis in the bottom row. With
$K_{\mbox{max}}$ = 1 (harmonic motions; top row, left panel), 
the algorithm converges very quickly to an
optimum, the same across all initializations of $A_0$ and $A_1$, but varying with
$\mu_n/\mu_s$. The maximum of $\phi$ is near $\pi/2$, the state of minimum drag, for all $\mu_n/\mu_s < 1$ except $2^{-0.5}$, close to isotropic friction. The minimum of $\phi$
increases monotonically with $\mu_n/\mu_s$. With $K_{\mbox{max}}$ increased to 3 (second panel,
top row), the optimal motions fall basically into two groups. All have peak values near $\pi/2$. Those
with $2^{-4} < \mu_n/\mu_s < 2^0$ (blue-green) have a closing phase that is more 
rapid than the opening phase, akin to the piecewise linear profiles studied previously.
 Those
with $2^{-7} < \mu_n/\mu_s < 2^{-4}$ (green-orange-yellow) have a plateau near
$\pi/2$, and an opening phase that is slightly more rapid than the closing phase. 
The minimum $\phi$ and the speeds of opening and closing vary slightly within each group.
With $K_{\mbox{max}}$ increased to 7 and 10 (third and fourth panels,
top row), the optima are similar, but with longer plateaus at $\pi/2$, and some motions
that decrease to $-\pi/2$. Following the previous discussion of negative $\phi$ values, such motions ``clap" together at $\phi = 0$ and then reopen to a state at $-\pi/2$ that is
the same as that at $\pi/2$. The corresponding values of 
$\lambda/\lambda_{\mbox{ub}}$ (normalized efficiency) are plotted in the middle row,
and generally increase in magnitude from $K_{\mbox{max}}$ = 1 (squares) to 3 (plusses), 7 (triangles), and 10 (crosses). At $\mu_n/\mu_s = 2^{-6}$ and $2^{-7}$, those with $K_{\mbox{max}}$ = 10 pass through $\phi = 0$ and they are less efficient than those with $K_{\mbox{max}}$ = 7. Thus, increasing the number
of modes can cause the iterates to converge to less efficient local optima. For other $\mu_n/\mu_s$ (except $2^{-6}$ and $2^{-7}$), there is good agreement of the best $\lambda/\lambda_{\mbox{ub}}$ values and the rescaled periods of the optimal motions
$(\mu_n\mu_s)^{1/4}T$ (bottom row) for $K_{\mbox{max}}$ = 3, 7, and 10, indicating convergence to
a particular motion. There are general trends towards higher efficiency and longer oscillation periods as the anistropy increases.

\begin{figure} [h]
           \begin{center}
           \begin{tabular}{c}
               \includegraphics[width=6in]{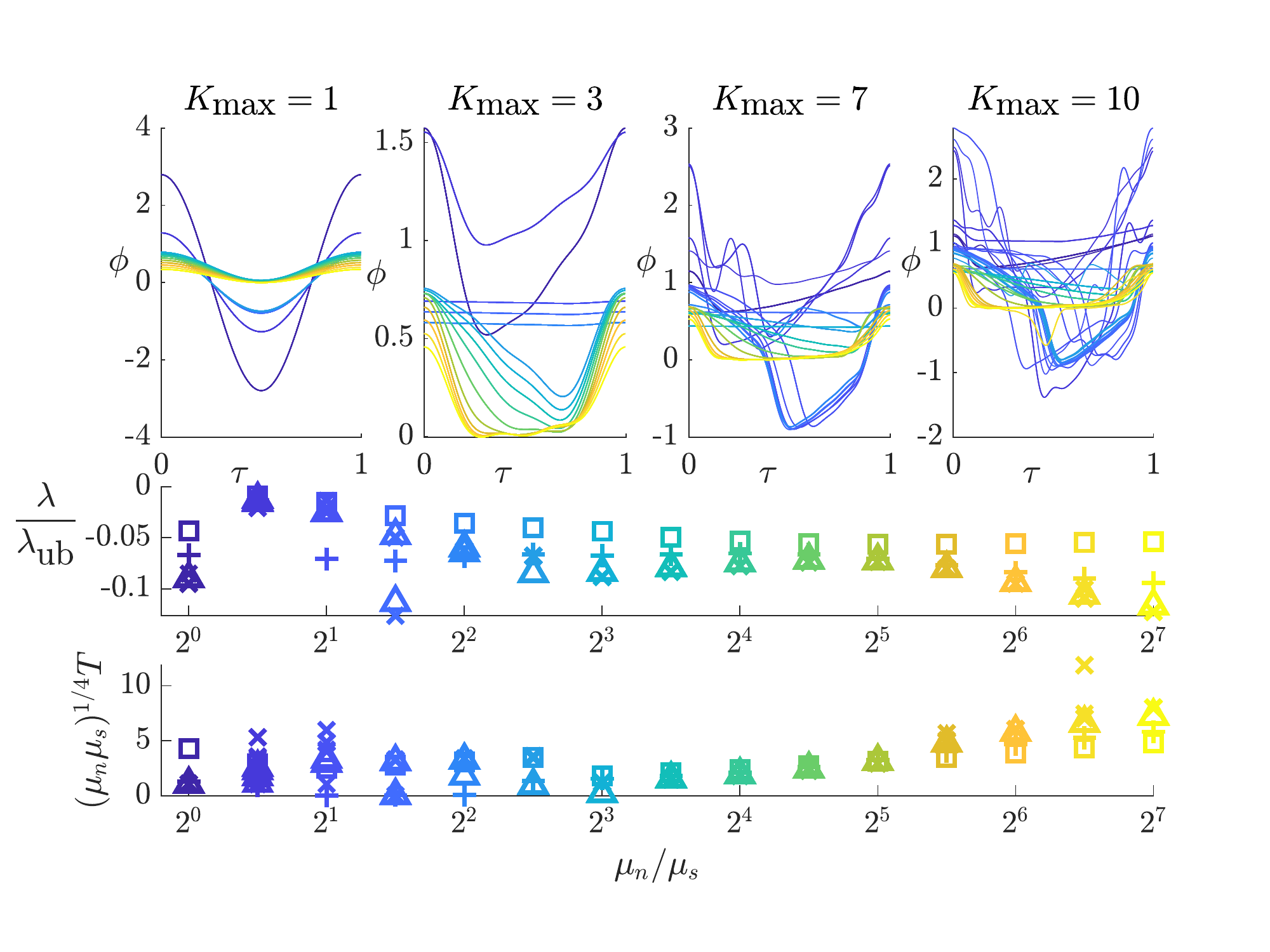} \\
           \vspace{-.25in}
           \end{tabular}
          \caption{\footnotesize Optimal motions in the space of truncated Fourier series when normal friction is dominant or friction is isotropic ($\mu_n/\mu_s \geq 1$). All other quantities are as described in the previous figure caption.
 \label{fig:OptMotionsHighMun}}
           \end{center}
         \vspace{-.10in}
        \end{figure}

Figure \ref{fig:OptMotionsHighMun} shows the same data for $\mu_n/\mu_s \geq 1$. Now
most of the optima are characterized by a minimum or plateau near $\phi = 0$ (the minimum
drag state), with slightly asymmetric opening and closing speeds, and maximum $\phi$ ranging
from 0.4 to 0.8. A wide range of other motions are found near isotropic friction (blue colors),
with $\phi$ more oscillatory and passing through zero in more cases. Nonetheless, many
of the optima at these $\mu_n/\mu_s$ are similar at $K_{\mbox{max}}$ = 7 and 10, leading to the clustering of $\lambda/\lambda_{\mbox{ub}}$ (middle row) and $(\mu_n\mu_s)^{1/4}T$
(bottom row) at certain values. There are again general trends towards higher efficiency and longer oscillation periods as the anisotropy increases, with more scattered behavior near the isotropic case.

Taken together, figures \ref{fig:OptMotionsLowMun} and 
\ref{fig:OptMotionsHighMun} show that across $\mu_n/\mu_s$,
and even at the highest $K_{\mbox{max}}$, many of the optimal
motions can be divided into monotonic opening and closing phases, and are thus similar
to the motions in the previous sections. The peak $|\lambda|/\lambda_{\mbox{ub}}$ reaches
0.55 when tangential friction is dominant, and about 0.12 when normal friction is dominant,
with gradients toward increasing $|\lambda|/\lambda_{\mbox{ub}}$ with further 
increases in anisotropy.

\section{Other power laws for the resistive force \label{sec:Fluid}}

So far we have considered the Coulomb model of sliding on a dry surface, 
modeled by a resistive force that is independent of velocity magnitude \cite{persson2000sliding,popov2010contact,GuMa2008a,HuNiScSh2009a,HuSh2012a,pennestri2016review,aguilar2016review}.
Other resistive force laws arise when solid bodies slide on or swim through fluids. Here resistance increases with velocity magnitude, as a power law in certain cases. These cases are described by a generalization of (\ref{frictiondelta})--(\ref{delta}):
\begin{align}
\mathbf{f}(s,t) &\equiv - \left(\mu_n \hat{\mathbf{n}} \hat{\mathbf{n}}^{\top} + \mu_s \hat{\mathbf{s}} \hat{\mathbf{s}}^{\top}\right) \mathbf{v}_p \label{frictionp} \\
\mathbf{v}_p &\equiv \left(\partial_t x, \partial_t y\right)^\top\left(\partial_t x^2 +\partial_t y^2 + \delta^2\right)^{(p-1)/2}, \label{vp}
\end{align}
\nn where $\mathbf{v}_p$ is a vector in the direction of the local body velocity, with magnitude proportional to speed to the power $p$ (for $\delta = 0$). In (\ref{vp}) we retain the notation $\mu_n$ and $\mu_s$ for the resistance coefficients, as well as the regularization
term $\delta \ll 1$ for consistency with the previous results, though its effect is even more negligible when
$p > 0$ (so the force already rises continuously from zero at zero velocity when $\delta = 0$). The
case $p = 1/2$ can describe a smooth solid object sliding on a wet substrate \cite{persson2000sliding}. The motion is resisted by viscous shear forces in a thin lubrication layer between the body and the substrate. The resistive force magnitude $f$ is proportional to the object's speed $v$ divided by
the depth of the lubrication layer $h$. The object's weight (assumed constant) is balanced by lubrication pressure $\sim v/h^2$, so
$h \sim \sqrt{v}$, and giving $f \sim v/\sqrt{v} = \sqrt{v}$.  The constants of proportionality depend on the object's weight, shape, 
size, and the fluid viscosity. Additional complexities arise when considering a deformable object with nonuniform velocities, very thin fluid layers of the order of the object's or substrate's roughness, surface tension, and/or non-Newtonian fluids \cite{persson2000sliding}.

Values of $p$ in the vicinity of 1 can describe microscopic bodies sliding on wet surfaces or swimming in viscous fluids (e.g.~Resistive Force Theory) \cite{lauga2009hydrodynamics,sauvage2011elasto,shen2012undulatory,rabets2014direct}. For microscopic bodies, the body inertia
(the left side of (\ref{Newton})) is negligible and a two-link body does not locomote \cite{purcell1977life,childress1981mechanics,lauga2009hydrodynamics}. However, we consider the case $p = 1$ to contrast this well-known case with the 
results at $p = 0$. Values of $p$ between 1 and 2 may occur in steady flows for
macroscopic flexible bodies \cite{Vogel1994,ASZ_Nature_2002,miller2012reconfiguration,lopez2014drag} and Blasius boundary layer drag on streamlined bodies \cite{prandtl1904flussigkeitsbewegung,blasius1908boundary}; and $p$ = 2 occurs in blade element \cite{blake1979mechanics,walker2000mechanical} and slender/elongated body \cite{candelier2011three,singh2012energy} models of swimming by macroscopic bodies. For macroscopic bodies immersed in fluids, fluid inertia introduces additional acceleration-dependent (reactive) forces, 
but here we limit the discussion to velocity-based (resistive) forces with powers $p$. 

In this section, we consider the extent to which the optimal motions of the previous section 
depend on the particular choice $p = 0$. With the force law (\ref{frictionp}), the average power or rate of work done against friction (\ref{power}) becomes

\begin{align}
P = \int_0^{1} 2 \int_{0}^{1/2} \left(\mu_n
\left(\partial_t\mathbf{X}\cdot \hat{\mathbf{n}} \right)^2
+ \mu_s \left( \partial_t\mathbf{X}\cdot \hat{\mathbf{s}} \right)^2\right)\left(\partial_t x^2 +\partial_t y^2 + \delta^2\right)^{(p-1)/2} \,ds \,d\tau. \label{powerp}
\end{align}

The efficiency (\ref{lambda}) is generalized to 
\begin{align}
\lambda \equiv \mbox{sign}(U)|U|^{p+1}/P. \label{lambdap}
\end{align}
\nn as $P$ now scales approximately as velocity to the $p+1$ power. Repeating the steps after equation
(\ref{Newtonu}), $u(\tau)$ now depends on $T$, $\mu_n$, and $\mu_s$ through the combinations $\mu_n/\mu_s$ and $(\mu_n\mu_s)^{1/(4-2p)}T$. The scaled average velocity and power are
$(\mu_n\mu_s)^{1/(2p-4)}U$ and $(\mu_n\mu_s)^{3/(2p-4)}P$ respectively. The scaled efficiency is still $\sqrt{\mu_n\mu_s}\lambda$, and the relative efficiency is still given by (\ref{relefficiency}).

\begin{figure} [h]
           \begin{center}
           \begin{tabular}{c}
               \includegraphics[width=5.5in]{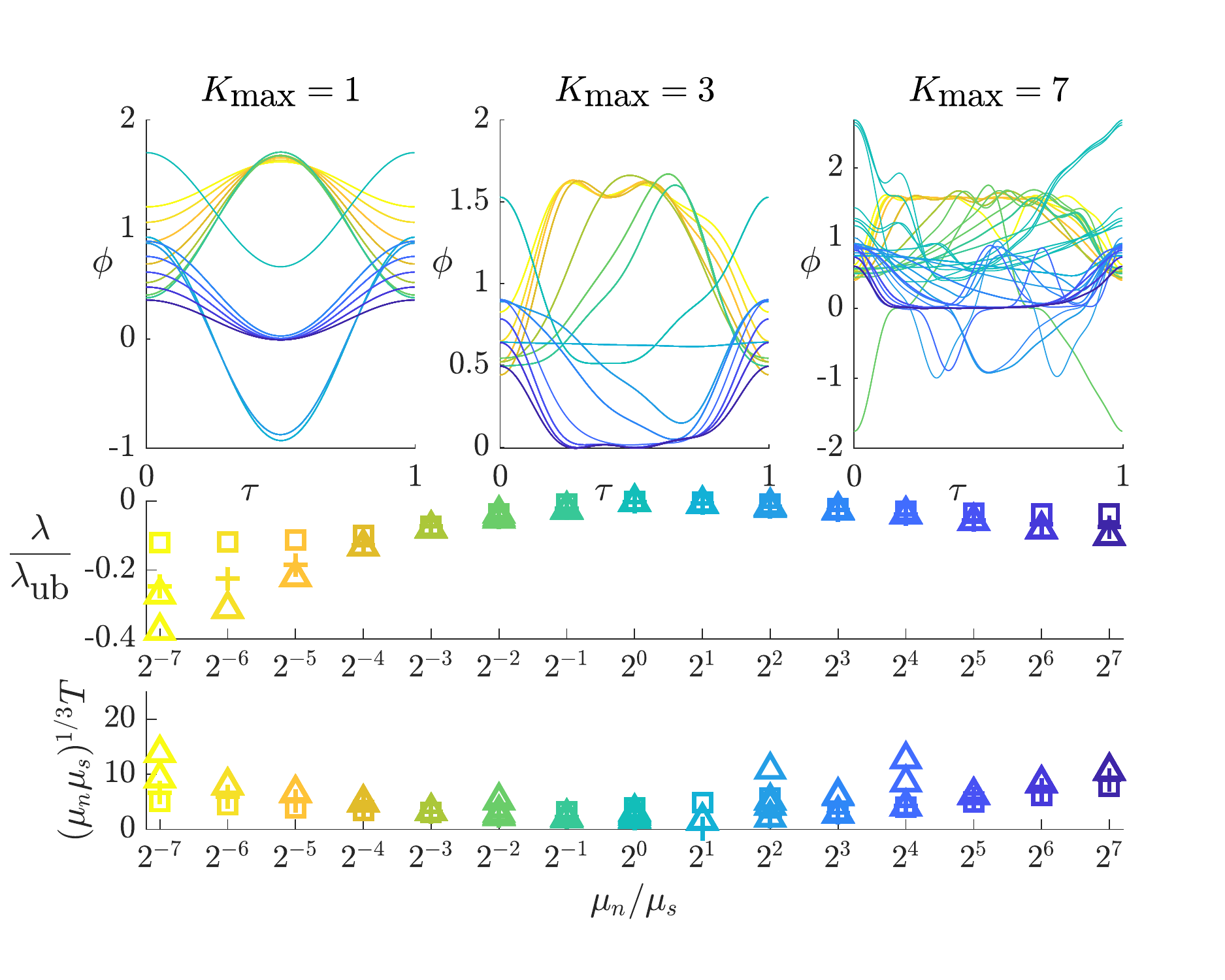} \\
           \vspace{-.25in}
           \end{tabular}
          \caption{\footnotesize Optimal motions in the space of truncated Fourier series with resistive force law exponent $p = 0.5$.
 \label{fig:OptMotionsp0p5}}
           \end{center}
         \vspace{-.10in}
        \end{figure}

Figure \ref{fig:OptMotionsp0p5} shows the optimal motions with $p = 0.5$. For simplicity,
we plot motions across $\mu_n/\mu_s$ in a single figure, and show results
for $K_{\mbox{max}}$ = 1, 3, and 7 only. Comparing optimal motions with those at $p = 0$ in the previous two figures (noting that a single color scale is used for both large
and small $\mu_n/\mu_s$ for $p = 0.5$), we find very good agreement. The stroke amplitudes are slightly
larger with $p = 0.5$, and the optima are most different in the vicinity of isotropic friction, where they show a diversity of functional forms. The relative efficiencies are somewhat lower with $p = 0.5$, particularly near isotropic friction, and the scaled periods are larger.

\begin{figure} [h]
           \begin{center}
           \begin{tabular}{c}
               \includegraphics[width=5.5in]{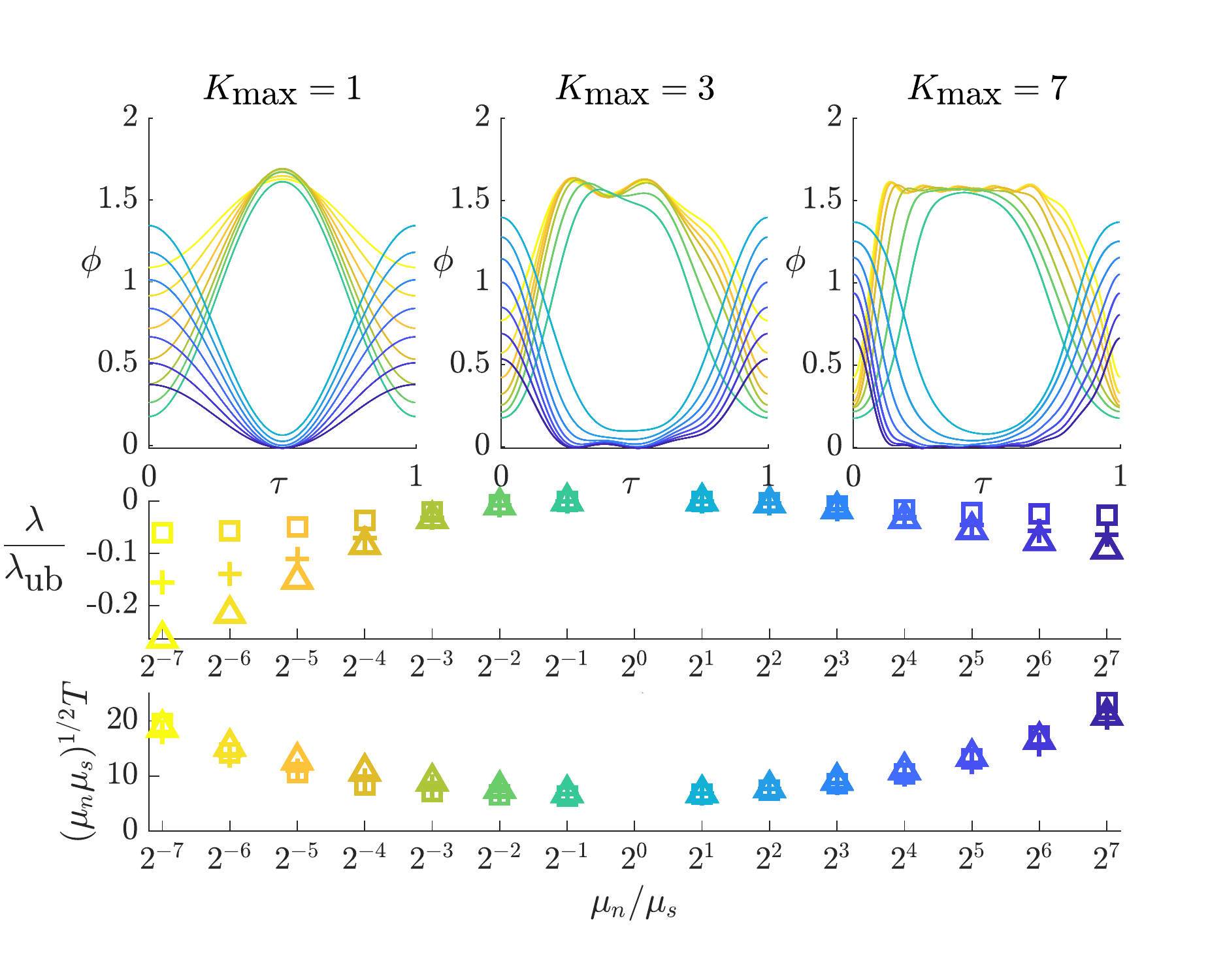} \\
           \vspace{-.25in}
           \end{tabular}
          \caption{\footnotesize Optimal motions in the space of truncated Fourier series with resistive force law exponent $p = 1$.
 \label{fig:OptMotionsp1}}
           \end{center}
         \vspace{-.10in}
        \end{figure}

The optimal motions with $p = 1$ are shown in figure \ref{fig:OptMotionsp1}. The motions are much
smoother and show much less variability, particularly near isotropic friction. There is only a single optimum for each $K_{\mbox{max}}$
and $\mu_n/\mu_s$, and $|\lambda|/\lambda_{\mbox{ub}}$ increases monotonically with
$K_{\mbox{max}}$. It can be shown analytically that $U$ = 0 with isotropic friction and $p = 1$,
so the relative efficiency (not plotted) is always zero when $\mu_n/\mu_s = 1$. 
Relative to $p = 0$ and 0.5, the optimal motions now have still larger amplitudes,
 smaller relative efficiencies, and larger scaled periods. If we assume $\mu_n$ and $\mu_s$ are both fixed
and $\leq 1$ (the typical case) as we increase $p$ from 0 to 1, then $(\mu_n\mu_s)^{1/(4-2p)}$ decreases,
so the increase in the optimal $(\mu_n\mu_s)^{1/(4-2p)}T$ corresponds to an even greater increase in $T$.

\begin{figure} [!h]
           \begin{center}
           \begin{tabular}{c}
               \includegraphics[width=5.5in]{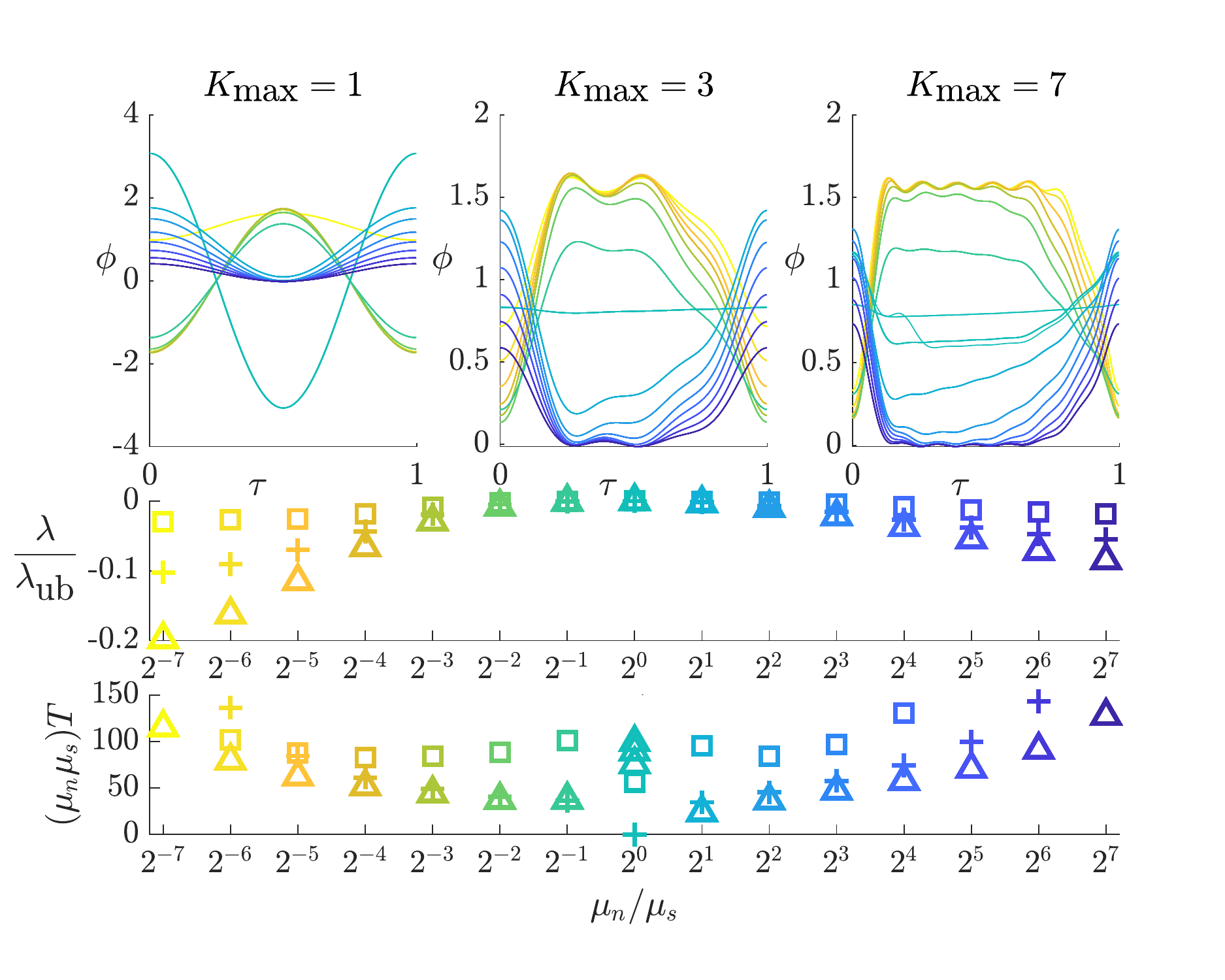} \\
           \vspace{-.25in}
           \end{tabular}
          \caption{\footnotesize Optimal motions in the space of truncated Fourier series with resistive force law exponent $p = 1.5$.
 \label{fig:OptMotionsp1p5}}
           \end{center}
         \vspace{-.10in}
        \end{figure}

Increasing $p$ further to 1.5, the optimal motions are shown in figure \ref{fig:OptMotionsp1p5}.
The optima still have relatively little variability of form, but have a slightly different form than those with $p$ = 1 for  $K_{\mbox{max}}$ = 3 and 7, particularly near isotropic friction ($2^{-3} < \mu_n/\mu_s < 2^3$). 
At $p = 1.5$, these motions have a longer plateau region, but the height of the plateau varies, and deviates from the minimum drag state (either $\phi = 0$ or $\pi/2$). At $p = 1$, by contrast, the motions always
get close to the minimum drag state but the graphs of $\phi(\tau)$ are more rounded, with less of a plateau (near isotropic friction). The general trend to larger scaled periods and smaller relative efficiencies with increasing $p$ continues. As $p$ is increased above 1.5, the scaled periods become much larger and eventually, for $p$ somewhat less than 2, convergence becomes more difficult to obtain because the Hessian matrix is ill-conditioned. A modified version of our algorithm may be able to handle this problem but we do not pursue it further here.


\section{Summary and conclusions}
We have studied the locomotion of a two-link sliding body, as an example where the speed of the body's movements, and the body's inertia, play an important role in sliding locomotion. The body's shape and its translational motion are each described by scalar functions of time, making it possible to solve for the motions throughout parameter spaces that capture essentially the full range of possible dynamics. Rotational motions are automatically avoided due to bilateral symmetry. By rescaling the solutions
we find that they depend on the speed of the motion and the geometric mean of the friction coefficients in
the single combination $(\mu_n\mu_s)^{1/4}T$, leaving the friction anisotropy $\mu_n/\mu_s$ as the most important physical parameter for describing the optimal motions. 

Efficient harmonic motions oscillate between the state of minimum drag (the body fully closed and horizontal, or fully open and vertical, depending on $\mu_n/\mu_s$) and another nearby state. The distance between the states changes monotonically with $\mu_n/\mu_s$ except near isotropic friction. The motions can be divided into power and recovery strokes, similar to those of jet-propelled swimmers when $\mu_n/\mu_s > 1$. 

We then considered two-phase (opening and closing) motions that are power laws on each phase.  Allowing
asymmetric speeds of opening and closing allows large increases in efficiency by allowing very rapid recovery
strokes, which result in less deceleration. Kinematic power laws that are different from unity allow efficiency improvements mainly for longer oscillation periods, and for very anisotropic friction ratios. These strokes
allow the body to ``coast" in its lowest drag state for much of the cycle, with rapid opening and closing motions.
The efficiency is sensitive to the functional form of the opening and closing motions. A range of other two-phase motions, some with locomotion in the direction opposite to those described, can also achieve high efficiencies and high speeds. We did not investigate these in detail.

In addition to motions with prescribed two-phase forms, we employed a local optimization algorithm to
determine efficient periodic motions more generally (as truncated Fourier series). In many cases,
the best local optima were similar to the two-phase motions already described, with ``coasting" in low drag states
for very anisotropic friction, and closer to the linear two-phase angle kinematics with more moderate anisotropy. A variety of other, more complicated motions were seen when friction is isotropic or slightly larger in the normal
direction. In general, relative efficiency can be much higher for the optima with $\mu_n/\mu_s < 1$  (e.g.~$> 60\%$ for $\mu_n/\mu_s = 2^{-8}$), because the body can move almost entirely normal to itself during the power
stroke and during coasting, using little energy against friction at those times. These results indicate that two link snake robots with smaller transverse than tangential friction could be competitive with more complicated designs in terms of efficiency.  

Finally, we examined the effect of changing the resistive force law to scale with velocity to powers 0.5, 1, and 1.5.
Increasing the power law increased the optimal scaled periods of the motions and decreased the relative efficiency of the optimal motions---particularly near isotropic friction. The Coulomb friction case was unusual for allowing isotropic motions with relative efficiencies comparable to the most efficient motions with $\mu_n/\mu_s \gg 1$.
With increasing resistive power law, the body angle kinematic functions became somewhat smoother, and with somewhat larger amplitudes, but many of the features, such as coasting in low-drag states, persisted.

These results indicate that the details of fast, intermittent kinematics are important
for efficient sliding locomotion of a two-link body. Some of the kinematics (such as burst-and-coast motions) may also improve efficiency when more complex spatial modes of deformation are allowed. Quasi-steady undulatory motions have been studied more often for sliding locomotion by snakes and snake robots, but including the effect of body inertia only enlarges the space of possibilities. For motions and maneuvers with rapid accelerations, such effects are unavoidable.

\begin{acknowledgments}
This research was supported by the NSF Mathematical Biology program under
award number DMS-1811889 (S.A.).
\end{acknowledgments}

\bibliographystyle{unsrt}
\bibliography{snake}

\end{document}